\documentclass[prl, nolongbibliography, twocolumn]{revtex4-2}

\usepackage{siunitx}% try to write down numbers and units in a consistent way
\usepackage{amsmath}
\usepackage{bm}
\usepackage{graphicx}
\usepackage{epsf}
\usepackage{xcolor}

\usepackage{color}
\begin{document}

\title{Persistent, controllable circulation of a polariton ring condensate}

\author{Q. Yao,$^1$ P. Comaron,$^2$ H. A. Alnatah,$^1$ J. Beaumariage,$^1$ S. Mukherjee,$^3$ K. West,$^4$ L. Pfeiffer,$^4$ K. Baldwin,$^4$  M. H. Szymańska,$^2$ and  D. W. Snoke$^1$}

\affiliation{$^1$Department of Physics and Astronomy, University of Pittsburgh 3941 O'Hara St., Pittsburgh, PA 15260, USA\\
$^2$Department of Physics and Astronomy, University College London, Gower St, London WC1E 6BT, United Kingdom\\
$^3$Joint Quantum Institute, University of Maryland and National Institute of Standards and Technology, College Park,
Maryland 20742, USA\\
$^4$Department of Electrical Engineering, Princeton University, Princeton, NJ 08544, USA
}

\begin{abstract} 
Persistent circulation is a canonical effect of superfluidity. In previous experiments, quantized circulation has been observed in polariton condensates, usually far from equilibrium, but persistent current in the absence of any stirring has not been seen. We report here the direct observation of persistent circulation of a polariton condensate with no driving force and with no observable change in time. We can cause the condensate to circulate in either direction on demand using a short laser pulse, after which the condensate continues to circulate for dozens to hundreds of rotations around a ring trap without any further stimulation. Our theoretical model successfully shows how the pulse, despite not carrying any angular momentum, causes the circulation.

\end{abstract}
\maketitle
%%%%%%%%%%%%%%%%%%%%%%%%%%  body  %%%%%%%%%%%%%%%%%%%%%%%%%%
\section{Introduction}
For any single-valued, well-behaved complex function, a closed loop in space changes phase by an integer multiple of $2\pi$, because after going around the loop the phase must match its value at the outset, {\em modulo} $2\pi$. This mathematical fact lies at the root of quantized circulation in superfluids and quantized flux in superconductors~\cite{introRef}. Quantized circulation has been seen in polariton condensates, including vortices in a highly nonequilibrium, turbulent polariton gas~\cite{deveaud, nphys_Marzena,panico2022onset}, small rings with short radiative lifetime~\cite{kav-ring}, and ``half-vortices'' both far from equilibrium~\cite{sanvitto} and in steady state~\cite{PNAS_liu}, with stochastic switching between circulation states. %in which the linear polarization of the condensate rotates by $\pm\pi$ around a closed path while the condensate phase also precesses by $\pm\pi$. 
A coherent standing wave in a ring with no circulation has also been seen~\cite{dreis, sturm}. Recently, vortex formation in ``rotating bucket'' experiments~\cite{lagou-arxiv, RotatingBucket}, which required active stirring, has been reported, as well as optical vortex switching~\cite{assman}. Vortices have also been observed in weakly-interacting optical systems~\cite{weiss}. 

However, no prior experiments have shown the effect of persistent quantized current, in the canonical sense, of a polariton superfluid. (The Supplementary Material gives a review of prior experiments.) This phenomenon relies not only on the fact that the condensate wave function is single-valued, but on the property of superfluids that they flow without friction. The canonical example is when a superfluid is given a brief kick to start rotating, after which it continues to circulate indefinitely without stirring. 

This means that a successful experiment to prove persistent circulation must demonstrate all of the following:
\begin{itemize}
\item It must measure circulation, namely, a net phase gradient around a closed path.
\item It must show that there is {\em no} circulation when the condensate is not stirred. Otherwise, the circulation could be explained by a built-in chirality of the underlying system (e.g., a ``toilet bowl'' configuration with chiral static profile in addition to a drain), which would be the equivalent of an ongoing stirring.% This rules out a large number of prior experiments that always saw circulation at the same location under the same conditions, without a controllable driving force to start the circulation. 
\item It must show that the driving force is fully ``off'' during the circulation. %A constant stirring that produces circulation is not an example of persistent circulation.
\item It must measure zero decay constant of the circulation. Circulation that persists past the driving force, but decays away on a finite time scale, is analogous to metal with small, but nonzero resistance, which cannot be claimed to be a superconductor.

\end{itemize}
We report here the observation of persistent circulation satisfying all of the above criteria. The spatially-resolved phase pattern of the condensate can be monitored non-destructively, {\em in situ}, in steady state, long after all transients have died down, and shows no decay of the circulation for as long as we can measure. A steady-state condensate is made possible by the long lifetime of the polaritons in our structures, which allows them to cool down to very low energy within their lifetime, as well as the high degree of uniformity of our ring trap.

%Earlier work showed quantized circulation with either a measurable decay time of the coherence, active stirring, or which always occurred and could not be turned off. 

 We also show deterministic control of the direction of circulation. It is a surprise that a single, non-resonant probe pulse can induce circulation, since it has negligible angular momentum on its own. However, we present a simplified theoretical model that shows how that this is possible because the probe gives a sideways kick in the ring to regions of slightly higher density.

\section{Experimental method and results}

The physics of polariton condensates in semiconductor microcavities is well established; for reviews see Refs.~\cite{caru,yama-deng,littlewood-BECbook, baum-BECbook}. Due to their photonic component, polaritons have very light mass ($\sim 10^{-4}$ times a vacuum electron mass), which allows condensation in the range of 5-50 K, while their excitonic component gives rise to relatively strong repulsive interactions.  Numerous  intrinsic nonequilibrium effects (e.g. Refs.~\cite{bloch-kpz, Keeling-KPZ, Szymanska-KPZ, Comaron_2021}) have been reported using polaritons with short lifetime (1-20 ps). For the past decade, however, it has been possible to create structures with polariton lifetimes of 300-400 ps, allowing macroscopic flow over hundreds of microns~\cite{nelsen,steger} and equilibration condensation, confirmed by fits of the particle occupation number as a function of energy to an equilibrium Bose-Einstein distribution~\cite{MITPRL, hassan, comaron2024coherence}. 
These long-lifetime structures have also been used to demonstrate the Bogoliubov branches of a polariton condensate~\cite{ostro}.

\begin{figure}
    \centering
    \includegraphics[width=.8\columnwidth]{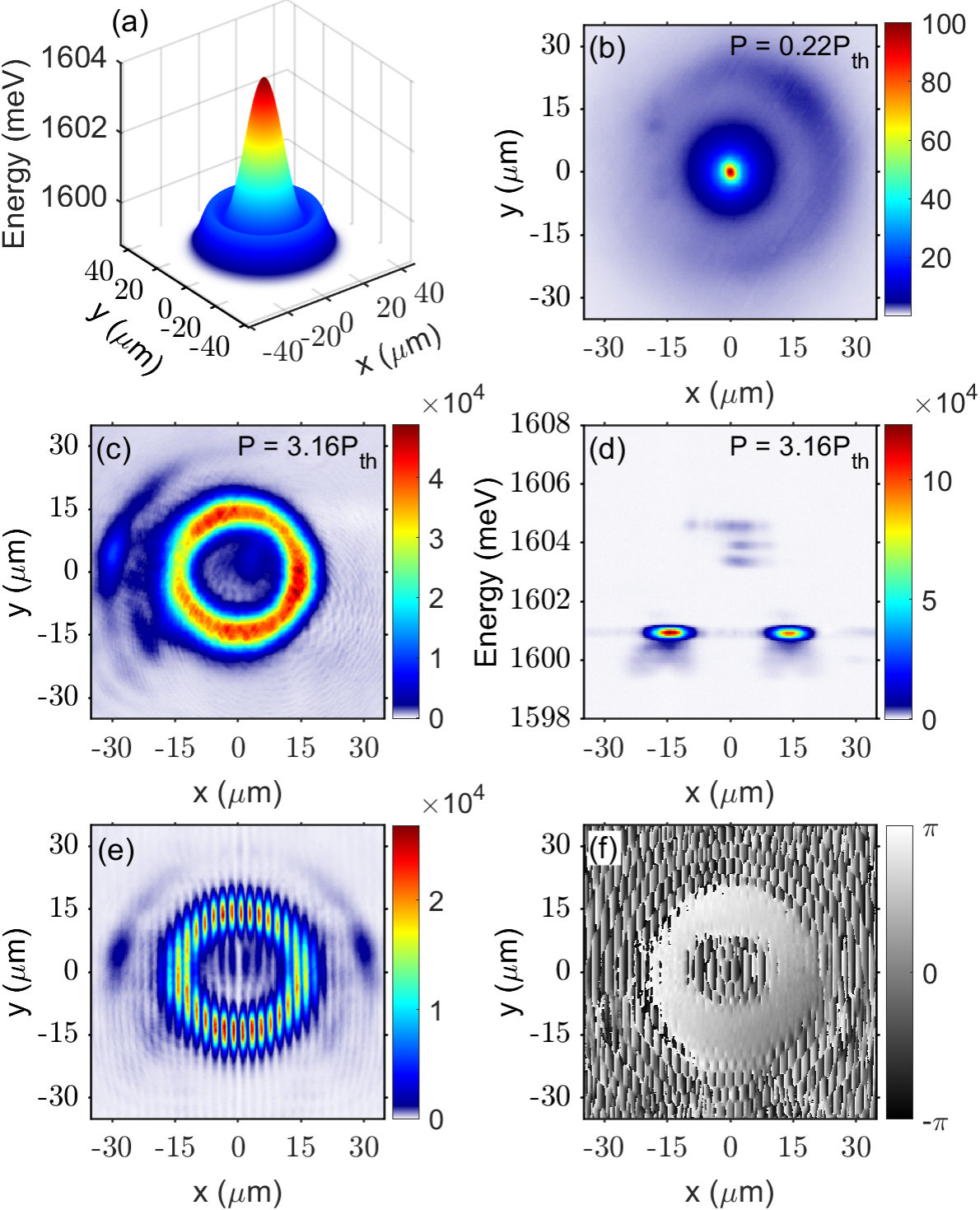}
    \caption{Characteristics of the ring condensate in the absence of circulation. (a) Approximate energy profile of the optically generated potential. (b) Experimental polariton PL below the threshold power ($P=0.22~P_{th}$). (c) Experimental polariton PL  above the threshold power ($P=3.16~P_{th}$). (d) Energy distribution along $x = 0~\mu m$ for (c). (e) Interference pattern between two images of the ring condensate, with one flipped across the $y$-axis. The experimental conditions were the same as (c). (f) Phase map extracted from the interference pattern in (e).}
    \label{fig:noCir}
\end{figure}

For the experiments described here, we used the same structure with long polariton lifetime as used previously (e.g., Refs. \cite{nelsen, steger, ostro}), namely an AlGaAs/GaAs microcavity with very high $Q$-factor (ca.~300,000). 
We generated the ring condensate by pumping the sample non-resonantly with a stabilized M Squared wavelength-tunable laser, using a spatial light modulator (SLM) to generate a ``Mexican hat'' potential energy profile, as shown in Figure~\ref{fig:noCir}(a). For these experiments we used a region of the sample in which the polaritons had approximately 50\% exciton fraction. The pump laser not only generated polaritons in the system but also formed a ring trap due to the repulsive interaction between excitons and polaritons. %Fig.~\ref{fig:noCir}(a) illustrates the optical trap generated by the pump laser below the critical density threshold $P_{th}$ for condensation (defined in the supplemental material). 
Chopping the pump laser with a 1.7\% duty cycle, with pulses of duration $41.6~\mu$s, was used to reduce total heating; as seen below, these pulses are very long compared to the dynamics. For more experimental details, see the Supplemental Material.

The light emission from the microcavity corresponds directly to the wave function of the polariton condensate, so that we can obtain both the intensity  and the phase information.
Below the threshold pump power, the PL was brightest at the pump location, as shown in Fig.~\ref{fig:noCir}(b). 
Above the threshold density for condensation, the intensity distribution underwent a dramatic shift as the condensate found the global minimum energy state, which in this case was a ring condensate in the circular trap (Fig. \ref{fig:noCir}(c)). The corresponding energy distribution along $x=0~\mu m$ (Fig. \ref{fig:noCir}(d)) shows a mono-energetic condensate. This is consistent with prior observations of polariton condensation using stress or etched trapping~\cite{davidPillar, PNAS_liu}. Unlike prior experiments~\cite{shouvikRing, PNAS_liu}, we created a trap which had almost zero gradient in the polariton energy by careful adjustment of the optical pump profile, so that the condensate did not feel any force which could cause it to flow towards any particular side.

We thus have a steady-state condensate with a well-defined wave function that has only a tiny fraction of its particles lost and regenerated at any point in time. This then becomes the ``system'' that we can excite with a probe and study using time-resolved time interference imaging.
Fig.~\ref{fig:noCir}(e) shows the interference pattern when two copies of the spatial image of the ring condensate PL are overlapped, with one of the images flipped $x\rightarrow -x$. The experimental condition was the same as Fig.~\ref{fig:noCir}(c). Fringes are seen from one side to the other, implying that the coherence extends across the whole ring. A typical measurement recorded 100 images with integration time 100 ms for each image. In the absence a short probe pulse, there were always an equal number of fringes at the top and at the bottom. This is the expected interference pattern when the ring condensate is not circulating. The phase map, Fig.~\ref{fig:noCir}(f), was extracted from the interference pattern using the method discussed in the Supplemental Material; it shows that the phase of the condensate fluctuates only by a small amount, with no net circulation.

\begin{figure}[t]
    \centering
    \includegraphics[width=.8\columnwidth]{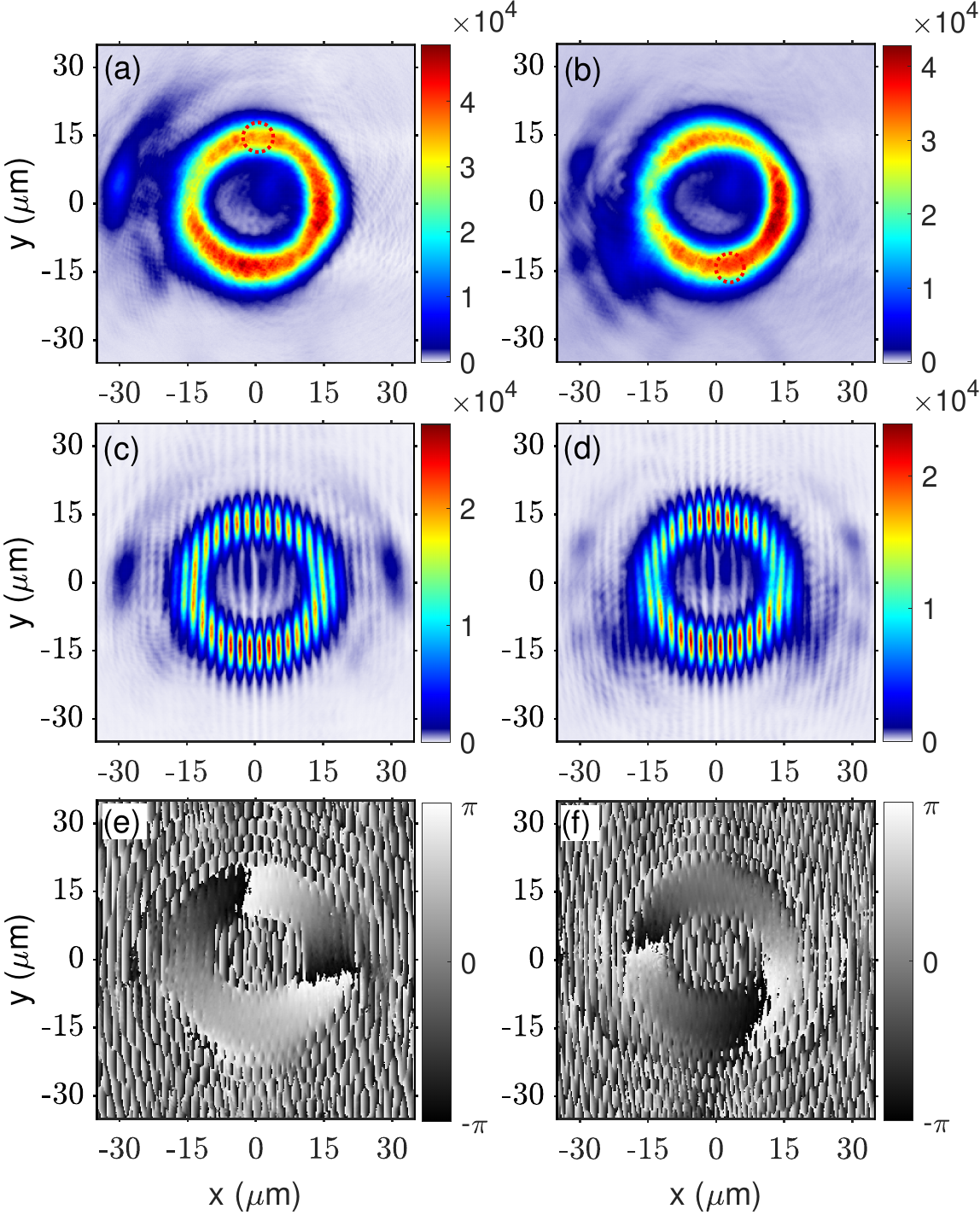}
    \caption{Experimental data of the circulating ring condensate. (a) Intensity distribution of the condensate with the probe at the top of the ring. The red dotted circle labels the location of the probe. (c) The time-averaged interference pattern. (e) Phase map extracted from (c). Images (b), (d) and (f) correspond to a case with probe pulse  moved to the bottom of the ring, as indicated by the dashed red circle in (b).}
    \label{fig:cirExp}
\end{figure}

To make the ring condensate circulate, we sent a short (ca.~1~ps) probe laser pulse onto the condensate at a localized region about 5~$\mu$m in diameter. This probe laser, like the pump laser, was non-resonant with central wavelength at $730$ nm. Fig.~\ref{fig:cirExp}(a) shows a typical intensity distribution of the circulating condensate when the probe was focused on the top side of the ring (indicated by the dotted red circle). Fig. \ref{fig:cirExp}(c) shows the time-averaged interference pattern in this case, recorded in the same way as Fig.~\ref{fig:noCir}(e).

Since the ring condensate circulates with no rotation of its electromagnetic polarization (see the Supplementary Material for measurements of the polarization), the phase change of the condensate around the ring must be $2\pi$N, where $N$ is an integer, with a phase defect (which may be called a ``vortex core'') at one point inside the ring. This implies an interference pattern with a different number of fringes on the top and at the bottom of the image. As seen in Fig.~\ref{fig:cirExp}(c), there are two more fringes at the bottom, which corresponds to $N=1$. 
Fig. \ref{fig:cirExp}(e) shows the phase map extracted from this interference pattern. A total of $4\pi$ winding is seen because the interference pattern gives the phase of the condensate relative to itself in the opposite direction.

When the stirring pulse was moved to the bottom of the ring, the circulation direction reversed. %while the intensity distribution didn't change much, as shown in Figs. \ref{fig:cirExp}(b),(d), and (f). 
As discussed in the Supplemental Material, we moved the location of the stirring pulse to various points about the ring and mapped the direction of the circulation induced. We found that the circulation was sensitive to the probe position when the probe moved in a region with nonuniform intensity, but the direction of circulation was deterministically controllable by the location of the stirring pulse, consistent with our numerical theoretical model, discussed below and in the Supplemental Material.

\begin{figure}[t]
    \centering
    \includegraphics[width=.8\columnwidth]{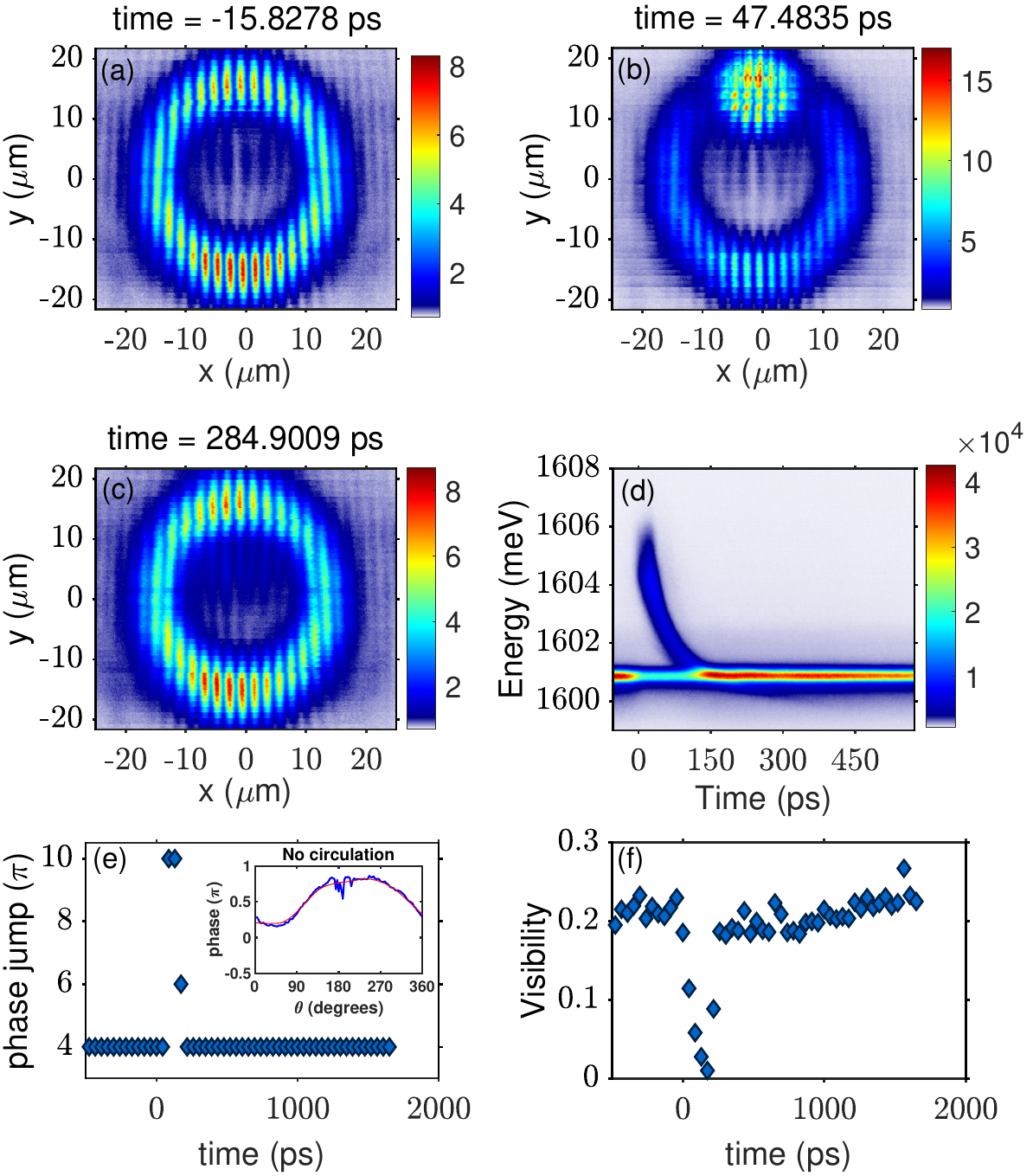}
    \caption{(a)-(c): Experimental interference patterns (a) before the probe pulse, (b) at the same time as the probe pulse, and (c) after the probe pulse. (d) Polariton energy as a function of time at the location of the probe. Note that the streak camera has a reverse trace that gives a constant-energy background signal corresponding to the condensate at much later time. (e) Phase jump versus time, when the probe pulse occurs at $t=0$ ps. Inset: phase as a function of angle around the ring for a non-circulating ring condensate, showing no $4\pi$ crossings. The blue curve is the extracted phase and the red curve is the smoothed data. (f) The visibility of the interference patterns on the sides of the ring, as a function of time, for the same conditions as (e).}
    \label{fig:phase}
\end{figure}

We used a streak camera with spatial imaging to see how the interference pattern changed in time. Fig.~\ref{fig:phase}(a)-(c) shows the snapshots at different times, where $t=0$ is defined as the time of maximum intensity of the probe pulse, for the case when the probe pulse is focused at the top of the ring, as in Fig.~\ref{fig:cirExp}(b). 
Since the probe pulse entered the system every 13.2 ns, corresponding to the 76 MHz repetiton rate of the pulsed laser, the signal at negative time includes the pattern remaining from the previous probe pulse. There is also an equal contribution from the reverse streak, corresponding to the signal 6.6~ns after the probe pulse. %\red{Is 6.6 correct? Half of 13.2?}
Figure~\ref{fig:phase}(d) shows the time-resolved spectrum at the location of the probe pulse. As seen in this figure, the probe pulse causes a large time-varying scalar potential variation, which reverts to the steady state conditions within about 300 ps. The contribution of the reverse trace of the streak camera can clearly be seen in Fig.~\ref{fig:phase}(d) as the part of the signal, which does not shift in energy during the transient occurring between $t=0$ and $t = 100$ ps.

The exciting result from these time-resolved measurements is that the interference pattern corresponding to circulation, with two extra fringes, persists for the entire period between the probe pulses with {\em zero} discernible degradation, which implies that the persistence time of the circulation is $\gg 14$ ns. Figure~\ref{fig:phase}(e) shows the extracted phase winding from the phase-winding images as a function of time, calculated as how many times the phase around the ring crosses zero. (In order to exclude the effect of the noise, the data was reasonably smoothed.) The insert of Fig.~\ref{fig:phase}(e) shows the phase around a non-circulating ring condensate, in which there is no zero-crossing. 

As seen in Figure \ref{fig:phase}(e), before and after the probe pulse, there is constant $4\pi$ phase winding along the ring. The probe pulse excites a transient state of the system, with several phase jumps due to the creation of the multiple vortices, which then settles down to a single, constant, quantized circulation. The transient state induced by the probe pulse is consistent with the numerical model discussed below and in the Supplemental material. 

Fig.~\ref{fig:phase}(f) shows the visibility of fringes at the sides of the ring, for the same data set used in Fig.~\ref{fig:phase}(e).
The visibility dropped when the probe pulse entered the system, but after this short transient, apart from fluctuations there was no overall degradation of the patterns after 14~ns.

% \begin{figure}[t]
%     \centering
%     \includegraphics[scale=0.65]{Figphase_a.pdf}
%       \includegraphics[scale=0.35]{figphase_b.pdf}
%     \caption{(a) Phase jump versus time, when the probe pulse occurs at $t=0$ ps. Inset: phase as a function of angle around the ring for a non-circulating ring condensate, showing no $4\pi$ crossings. The blue curve is the extracted phase and the red curve is the smoothed data. (b) Theoretical image of what would be seen in the interference pattern if there were stochastic flipping of the direction of circulation on time scales short compared to the image integration time.}
%     \label{fig:phase}
% \end{figure}

We can compute the speed of the condensate flow as 
\begin{equation}
v =\frac{\hbar}{m}\nabla \theta = \frac{\hbar}{m}\frac{2\pi}{2\pi R},
\end{equation}
where $R$ is the radius of the ring. This gives $v \simeq 0.1~\mu$m$/$ps, and so a circulation time of $T = 2\pi R/v \simeq 800$ ps, implying 
that between probe pulses, which come every 13.2 ns, the condensate circulates around the ring about 16 times with no probe pulse present. Comparing to the lifetime of any single particle in the condensate of about 300 ps, we conclude that the condensate has a robust phase memory that is not destroyed by the continuous small loss and generation of particles in steady state.

%Although our data is taken as the average over many shots of the probe pulse, the data we obtain is consistent with the direction of circulation being the same every time. Figure~\ref{fig:phase}(b) shows the image that is generated by summing 50\% of one circulation and 50\% of the opposite circulation. As seen in this image, stochastic flipping of circulation direction that is averaged over would lead to blurring of the image, which is not seen in our data. \\

\section{Theoretical analysis}
The fact that the probe pulse induces circulation is initially surprising, because it is not resonant with the condensate and therefore does not directly impart linear or angular momentum to the condensate. Like the pump laser, it generates incoherent electrons and holes that turn into excitons and eventually into polaritons; this is well modeled by a time-varying scalar potential for the polaritons plus a source term for excitons; any net momentum of the generated hot electrons and holes is quickly damped out by phonon emission and absorption. This was verified experimentally by varying the angle of incidence of the probe laser, which changes the in-plane linear momentum imparted to the carriers; no effect from the angle of incidence could be observed.

To gain insight into how the probe pulse induces circulation,
we performed simulations by numerically solving the time-dependent  equations for polariton motion within the Truncated-Wigner (TW) approximation~\cite{caru}, coupled to the rate equation for the excitonic reservoir:
\begin{align}
 i \hbar d \psi  &= 
 \begin{aligned}[t] 
  dt\bigg[ 
\left( i \beta - 1 \right) \frac{\hbar^2  \nabla^2}{2 m } 
+ g_c|{\psi} |_{{\mathcal{W}}}^2 +g_R n_R + 
\\
 + \frac{i \hbar }{2} \left( R n_R  - \gamma_c \right) 
\bigg] \psi  + i \hbar dW_c 
\label{eq:GPE_pol} 
\end{aligned}
\\ 
\frac{d}{dt} n_R  &= 
\begin{aligned}[t]
P_r (\textbf{{r}}) - \left( \gamma_R + R |{\psi} |_{{\mathcal{W}}}^2 \right) n_R 
\label{eq:GPE_res}
\end{aligned}
\end{align}
where $\psi(t) \equiv \psi(\textbf{r},t)$ is the polariton field, $n_{R}(t) \equiv n_{R}(\textbf{r},t)$ the excitonic reservoir density and $m$ the polariton mass.
In Equations (\ref{eq:GPE_pol}) and (\ref{eq:GPE_res}), $\gamma_c$ and $\gamma_R$ define the decay rates of condensed polaritons and the excitonic reservoir respectively.
As required by the TW picture, in \eqref{eq:GPE_pol} the Wigner noise accounts for quantum fluctuations and is space- and time-correlated as $ \left< dW_c(\textbf{r},t)  dW_c(\textbf{r}^\prime,t) ~\right> =0$, $\left< dW_c(\textbf{r},t)  dW_c^*(\textbf{r}^\prime,t)  \right> = (\gamma_\mathrm{c} + R n_\mathrm{R}({\textbf{r}}))/2 \  \delta_{\textbf{r},\textbf{r}^\prime}dt$.

The renormalized density $|{\psi}|^2_{\mathcal{W}} \equiv
\left|{\psi} \right|^2 - {1}/{2a^2} $  includes the subtraction of the Wigner commutator contribution, where $a$ is the numerical 2D grid lattice spacing~\cite{comaron2018}.
The relaxation parameter $\beta$ sets the amount of energy relaxation and increases as the detuning $\delta = E_c-E_{ex}$ increases~\cite{estrecho2018single}.
The constants $g_c$ and $g_R$ are the strengths of polariton-polariton and polariton-reservoir interactions respectively; they can be estimated as $g_c = g_{ex}|X|^4$, $g_R = g_{ex}|X|^2$, with $g_{ex}$ the exciton-exciton interaction, and $X$ is the Hopfield coefficient~\cite{yama-deng}. The parameter $R = R_0 g_c/g_R$ quantifies the stimulated scattering rate of the reservoir excitons into the polariton condensate.

We introduced a Mexican-hat 2D pump profile using the experimental parameters, modelled by $P_r(\textbf{r})=P ({\gamma_c}/{R})[  \exp(-\left( {(\textbf{r}^2/R_{ring}^2 -1)}/{2 \sigma_{ring}^2 } \right)^2) + H \exp(-\left( {\textbf{r}^2}/{2 \sigma_{gauss}^2 } \right)^2)  ]$, with $H$ the intensity ratio between the central Gaussian and the ring functions.
The steady-state wavefunction is found by solving Eqs.~(\ref{eq:GPE_pol}) and (\ref{eq:GPE_res}) for the time evolution. 
We initiated the system evolution from a random configuration of $\psi(\textbf{r})$.
After some relaxation dynamics, dominated by the polariton-reservoir repulsive interaction, the steady state of both reservoir and polariton particles was fully reached. Depending on the initial noise seed chosen, the steady-state can have different spatial profiles, presenting regions of low and high density. This is due to the hole-burning effect~\cite{estrecho2018single} introduced by the repulsive exciton-polariton interactions.
\begin{figure}
    \centering
    \includegraphics[width=.8\columnwidth]{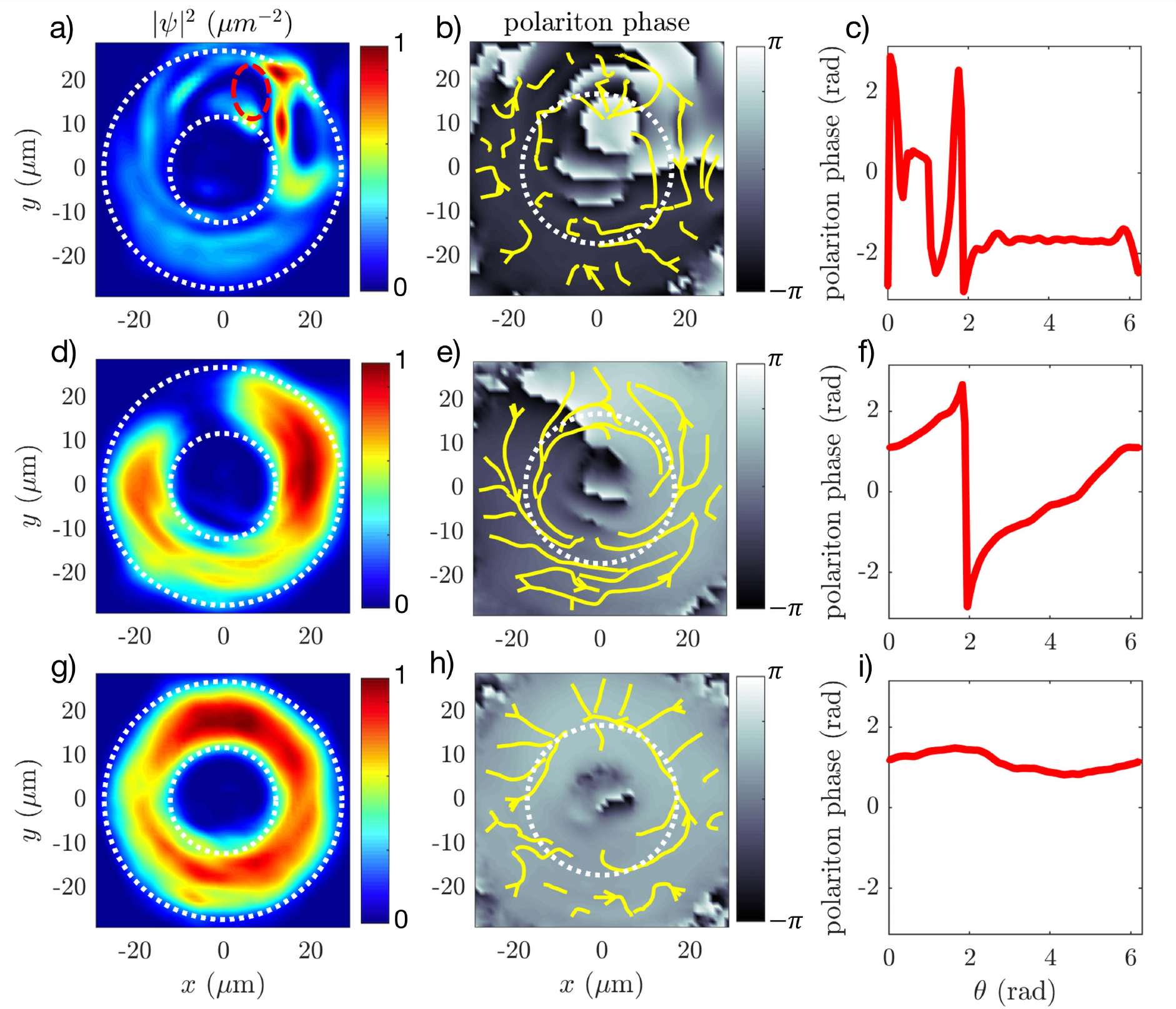}
    \caption{Numerical analysis. (a)-(b):  normalized density and phase of the condensate just after a spatially asymmetric probe pulse (red dashed circle in (a), $t=0.025$ ns). 
    Yellow lines  show the streamlines (curves tangent to the local velocity vector) of the polariton fluid. (c) Phase along the dashed white line in (b). Multiple $2\pi$ jumps reveal the presence of topological excitations. (d)-(f): the same at  $t=0.3$ ns. 
    A single $2\pi$ winding of the phase corresponds to a counter-clockwise current.  (g)-(i): the same at $t = 0.3$ ns when a symmetric probe is used. 
    }
    \label{fig:simCir}
\end{figure}
To model the external probe, at a time $t=0$ we added an extra exciton-reservoir density $n_R^\mathrm{probe} = H_{R} \mathcal{P} (1+\chi dW(t))$ with intensity $H_{R}$, where the noise $dW(t)$ is a zero-mean, classical Gaussian noise, and $\mathcal{P}$ is the spatial profile of the probe. The parameters used in the model are presented in the Supplementary Material.

Our numerical investigation reveals that a polariton current can be induced in the ring by an asymmetric probe focused on a minimum of the polariton density, in a manner consistent with the experiments. Essentially, the control over the direction of circulation depends on just two properties of the focal spot of the pulsed laser. One is the slight asymmetry of the focal spot, which gives it an oblong shape that effectively points in one direction. The other is where the spot is focused, relative to a region of high density of the condensate. Although the condensate has a single energy due to the condensate finding a common chemical potential across its whole spatial extent (see Fig. \ref{fig:noCir}(c)), it has density variations due to slight disorder of the underlying solid state structure and instabilities in the exciton cloud.  When the short laser pulse spot is to the left of a higher-density region, it pushes the condensate there to the right, while if it is to the right, it pushes the condensate to the left. 

Fig.~\ref{fig:simCir} shows typical numerical results. 
Inspection of the densities and phase dynamics shows that the perturbation introduced by the probe is responsible for creating topological excitations in the form of vortex-antivortex pairs (Figs.~\ref{fig:simCir}(a)-(c)). 
After this pair is created, the two vortices decouple and move in the fluid independently. By tracking the vortex-pair dynamics, we found that while most pairs recombine, one vortex moves to and eventually stops near the center of the ring trap, and another (opposite-circulation) vortex moves outside the ring condensate.
The decoupling of the vortex-antivortex pair creates a $2\pi$-jump of the phase, which gives rise to a persistent current inside the ring.
Such a current lasts as long as the vortex pair survives. 
In the simulations, we did not include static disorder outside the ring, and therefore the outer vortex eventually migrates towards the ring center, where it annihilates with its vortex twin inside the ring, and as a consequence the polariton circulation ends. However, if this outer vortex were to become pinned at a defect, which we believe happens in our experiment, consistent with earlier observations  ~\cite{deveaud}, then the circulation would continue indefinitely, as is observed here.
We found that the effect of circulation induced by the probe is robust to a wide variety of conditions, but requires the probe to be at least slightly asymmetric. 
An example of a symmetric probe focused on a minimum of the polariton density is shown in Fig.~\ref{fig:simCir}, panels (g)-(i). More details about the simulation can be found in Supplemental Material Section S4.

\section{Conclusions}
Our observation of the canonical effect of persistent circulation shows that the polariton condensate is indeed a superfluid by the standard definition. 
Our results sharply contrast with previous reports on persistent circulation in driven-dissipative systems, such as steady-state currents manifested under the influence of continuous external driving forces\cite{si15},\cite{si16}, or circulation fluctuated stochastically and did not remain non-circulating in the absence of a driving pulse \cite{PNAS_liu},\cite{si11},\cite{si17}. 
To the best of our knowledge, the experiments detailed herein are the most persistent steady-state circulation seen in quantum light fluids, characterized by the absence of measurable dissipation, with no contribution from disorder-induced chirality or fluctuations away from stable circulation.
Crucially, our current remains absolutely stable on the nanosecond timescale ---far exceeding any previously observed effect and marking a definitive step forward in the realization of persistent polariton superflows.
%

% Marzena, Paolo: remove the following sentence 
% This is made possible at large by the fact that in our experiments the polaritons have long lifetime, so that the system can reach near-equilibrium properties.
%&
Although it may at first seem surprising that we can generate circulation in either direction on demand using a single probe pulse that does not directly impart a net momentum to the condensate, our simulations show that this is indeed expected. The asymmetry of the probe-condensate system launches a vortex-antivortex pair with the two vortices moving in opposite directions, one toward the center and the other to the outside of the ring. Although random motion might bring these back together again, at which point they would annihilate, the outer vortex can, for example, become pinned at a defect or drift away from the area of interest, leading to the circulation of the condensate, which is for all intents and purposes permanent until the condensate is reset by another probe pulse or is ended.

The circulation seen here is a type of resettable optical memory that stores phase information, with a reset time of a few picoseconds and an essentially infinite storage time. An interesting future experiment would be to perform a single-shot experiment with just one probe pulse and track the persistence of the circulation we see well beyond the 14 ns limit of our present experimental setup. Given the nearly zero decay we see in the present experiments, we expect the circulation to last at least 100 times longer than 14 ns, i.e., microseconds or greater.

{\bf Acknowledgment} This work has been supported by the National Science Foundation (Grant No. DMR-2306977). M.H.S. and P.C. gratefully acknowledge financial support from EPSRC (Grant No. EP/V026496/1 and EP/S019669/1). PC was also supported from the EPSRC Grant No. EP/S021582/1.
This research made use of the HPC Computing service at University College of London. We thank F. Ducluzeau and F. Sidoli for computational assistance.

\bibliography{ref}

\end{document}

\title{Persistent, controllable circulation of a polariton ring condensate: supplemental document}

\author{Q. Yao,$^1$ P. Comaron,$^2$ H. A. Alnatah,$^1$ J. Beaumariage,$^1$ S. Mukherjee,$^3$ K. West,$^4$ L. Pfeiffer,$^4$ K. Baldwin,$^4$ M. Szymańska,$^2$ and  D. W. Snoke$^1$}

\affiliation{$^1$Department of Physics and Astronomy, University of Pittsburgh\\ 3941 O'Hara St., Pittsburgh, PA 15260, USA\\
$^2$Department of Physics and Astronomy, University College London\\ University College, Gower St, London WC1E 6BT, United Kingdom\\
$^3$Joint Quantum Institute, University of Maryland and National Institute of Standards and Technology, College Park,
Maryland 20742, USA\\
$^4$Department of Electrical Engineering, Princeton University, Princeton, NJ 08544, USA
}

\maketitle

\section{Experimental methods}
The microcavity sample was grown on a GaAs (001) substrate by molecular beam epitaxy (MBE). It consists of a $3\lambda/2$ microcavity and three sets of four GaAs quantum wells placed at the antinodes of the cavity photon mode. Two AlGaAs/AlAs distributed Bragg reflectors (DBRs) were used to create the microcavity, with 32 periods in the top DBR and 40 periods in the bottom DBR. An individual piece of the wafer was used for this experiment. The gradient of the sample was about 0.0031meV/$\mu$m, and the energy of the ring condensate was the same at different locations. The detuning of the ring is $1.1$~meV, corresponding to 47.54\% photonic fraction. The Rabi splitting was $\sim15.08$~meV according to the characterization results. More information about the sample can be found in Ref. ~\cite{samplePRB}
\begin{figure}[h]
    \centering
    \includegraphics[width=0.9\textwidth]{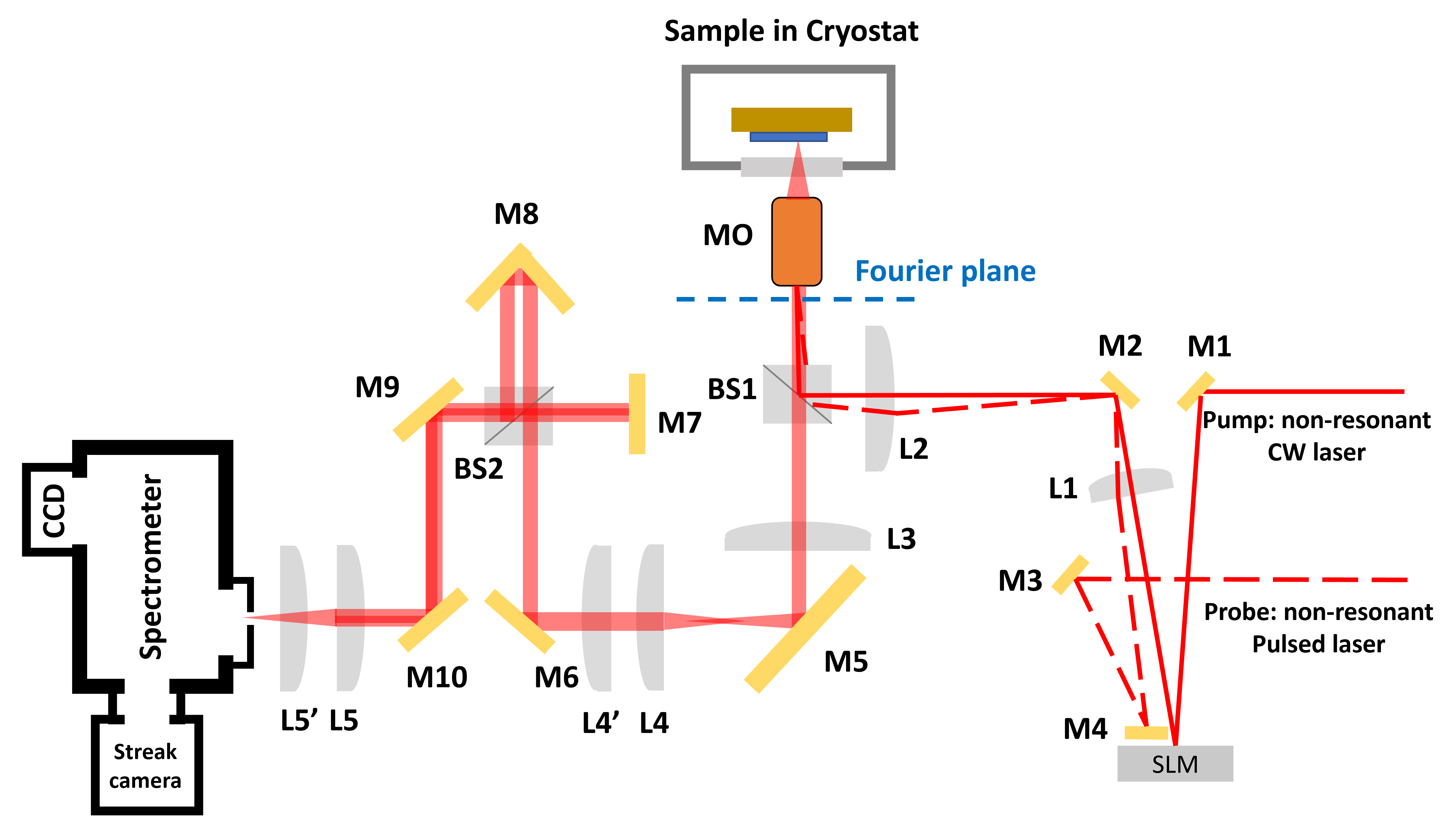}
    \caption{Schematic of the experimental setup. MO: microscope objective; BS: beamsplitter; SLM: spatial light modulator; L: lens; M (except M8): flat mirror; M8: retroreflector.}
    \label{fig:setup}
\end{figure}

During the experiment, a continuous-wave Ti:Sapphire laser, actively locked at $E_{pump} = 1698.6$~meV ($\lambda = 730$~nm), was sent to spatial-light modulator (SLM) to form a Mexican hat shape, and non-resonantly pumped the sample through a 50$\times$0.75-N.A. microscope objective. A mechanical chopper with a 1.7\% duty cycle at 400 Hz was placed in the pump laser path to reduce heating. All reported powers are the peak power of each short pulse. A Ti:Sapphire mode-lock laser with central wavelength at 730~nm and repetition rate at 76MHz was focused on the sample as the probe. The probe spot was not symmetric, with $\sim4.43~\mu$m FWHM in the horizontal direction, and $\sim3.01~\mu$m FWHM in the vertical direction.The same microscope objective was used to capture photoluminescence (PL) from the sample of 1599.4~meV (775.3~nm). The reflected laser was blocked by an energy filter (750~nm long-pass filter). Both real-space images and energy-resolved images were taken with a charged-coupled device (CCD) camera. We also obtained angle-resolved imaging by switching the lens combination. All measurements were performed by cooling the microcavity to low temperature (below 10 K) in a continuous-flow cold-finger cryostat. 

Fig.~\ref{fig:setup} is a schematic of the experimental setup. The cw laser beam (solid red line) is modulated by SLM. The pulsed laser beam (dashed red line) is close to the cw laser beam and incident at an angle. The thick red line represents the photoluminescence (PL). The blue dashed line labels the Fourier plane. Lens L4, L4$'$, L5 and L5$'$ are on flip mount. L4 and L5 are used for real-space imaging and L4$'$ and L5$'$ are used for momentum-space imaging. Mirror $\text{M6}$ and $\text{M10}$ are on flip mounts as well. When $\text{M6}$ and $\text{M10}$ are flipped down, the camera could take the real-space image of the condensate. When the mirrors are flipped up, the PL is sent to the interferometer, which has retroreflector $\text{M}8$ as one of the end mirrors to flip the PL along the symmetry axis. We can control a mirror in the spectrometer by the computer to send the light to the CCD camera or the streak camera.

\section{Numerical method of retrieving the phase from the interference pattern}
In this section, we first set up a model to show how the circulation would cause different number of fringes in the interference pattern and the method to extract phase map.  

\begin{figure}[h]
    \centering
    \includegraphics[scale = 0.5]{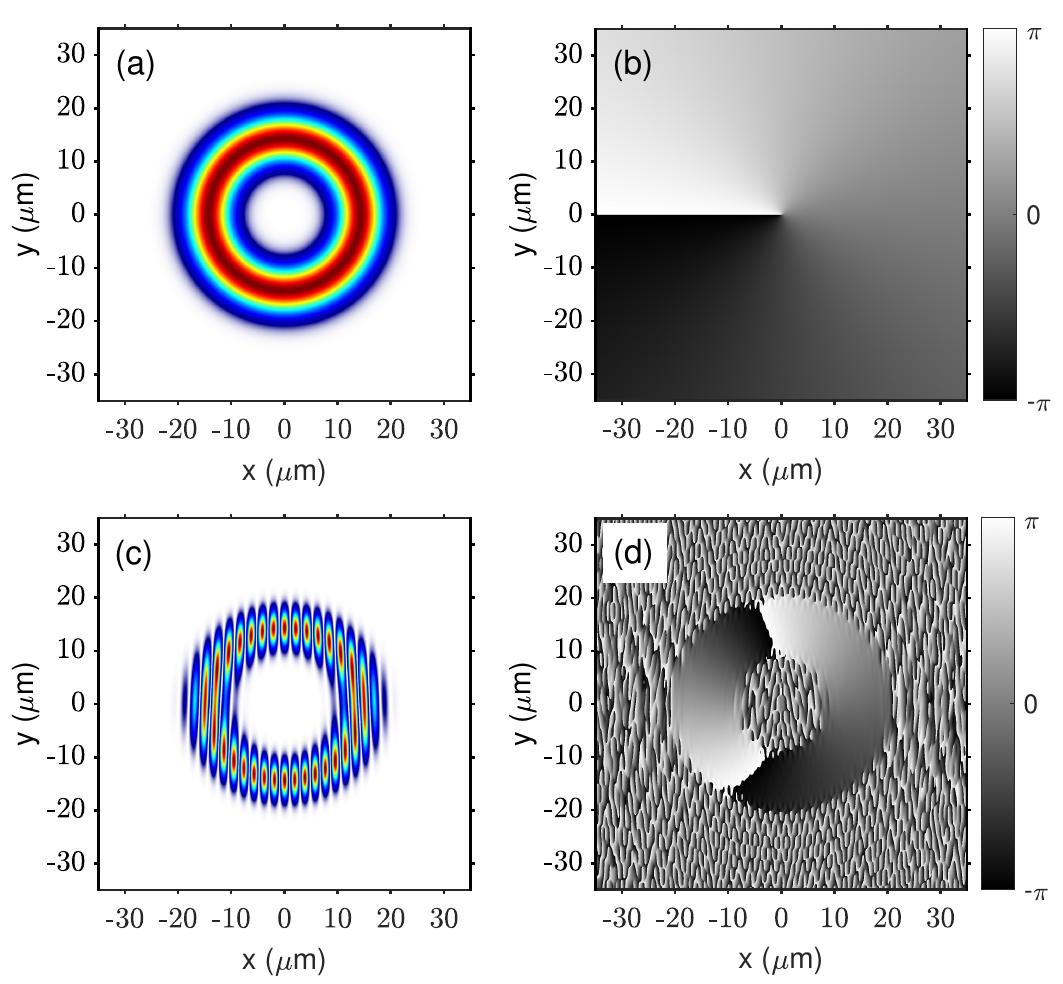}
    \caption{Simulation of a circulating ring condensate. (a) Intensity distribution. (b) Phase winding assigned to the condensate. (c) Predicted interference pattern for this condensate. (d) Phase map extracted from (c).}
    \label{fig:cirSim}
\end{figure}

The wavefunction of the ring condensate with circulation can be written as:
\begin{equation}
    \psi(r) = A(r)\exp{[i\phi(r)]}
\end{equation}
where the intensity distribution, $|\psi(r)|^2 = |A(r)|^2$, has a ring shape, and the phase, $\phi(r)$, changes from $-\pi$ to $\pi$, indicating the circulation with winding number 1. Fig.~\ref{fig:cirSim}(a) and (b) show the intensity distribution and the phase assignment respectively.

After the PL is sent to the interferometer, one copy carries an extra phase, $\hat{k}\cdot\hat{x} = k_x x \cos(\theta)$, and the phase change of the other copy is $-k_x x \cos(\theta)$, where $\theta$ is the incident angle. We assume the second copy is reflected by the retroreflector, which changes the original phase of the wavefunction $\phi$ to $\pi-\phi$. As a result, the interference pattern can be written as
\begin{equation}\label{eq:interference}
\begin{split}
   I(r) & = |\psi(r)\exp{[ik_{x}x\cos{\theta}]}+\psi^{*}(r)\exp{[i\pi-ik_{x}x\cos{\theta}]}|^2 \\
        & = A^2(r)\big(2-e^{i(2\alpha)}-e^{i(-2\alpha)}\big) = A^2(r)\big(2-2\cos{2\alpha}\big)
\end{split}
\end{equation}
where $\alpha = k_{x}x\cos{\theta}+\phi(r)$. Fig.~\ref{fig:cirSim}(c) shows the interference pattern, and there are two more fringes at the bottom than at the top, consistent with the experimental observation.

In order to extract the phase map from the interference pattern, we followed the method described in Ref.~\cite{si10}. We first transform the interference pattern to a reciprocal space map through 2D fast Fourier transform. From Eq.~\ref{eq:interference}, there are three components centered at $k=\pm 2k_{x}\cos{\theta}\,\text{and}\,0$ in the reciprocal space map. Both high-frequency components carry the phase information, $2\phi$. Then we only keep one of the high-frequency components, and shift it to the center of the reciprocal map, so that only the phase information is reserved. Finally, the 2D inverse fast Fourier transform is applied to reconstruct the phase map. Fig.~\ref{fig:cirSim}(d) is the reconstructed phase map. It has the same circulation direction as Fig.~\ref{fig:cirSim}(b) but has one more phase change. This is because the high-frequency components of the interference pattern is the phase difference between the condensate and itself in the opposite direction. 

\newpage

\section{Comparison between no circulation and stochastic flipping}

\begin{figure}[h]
    \centering
    \includegraphics[scale = 0.4]{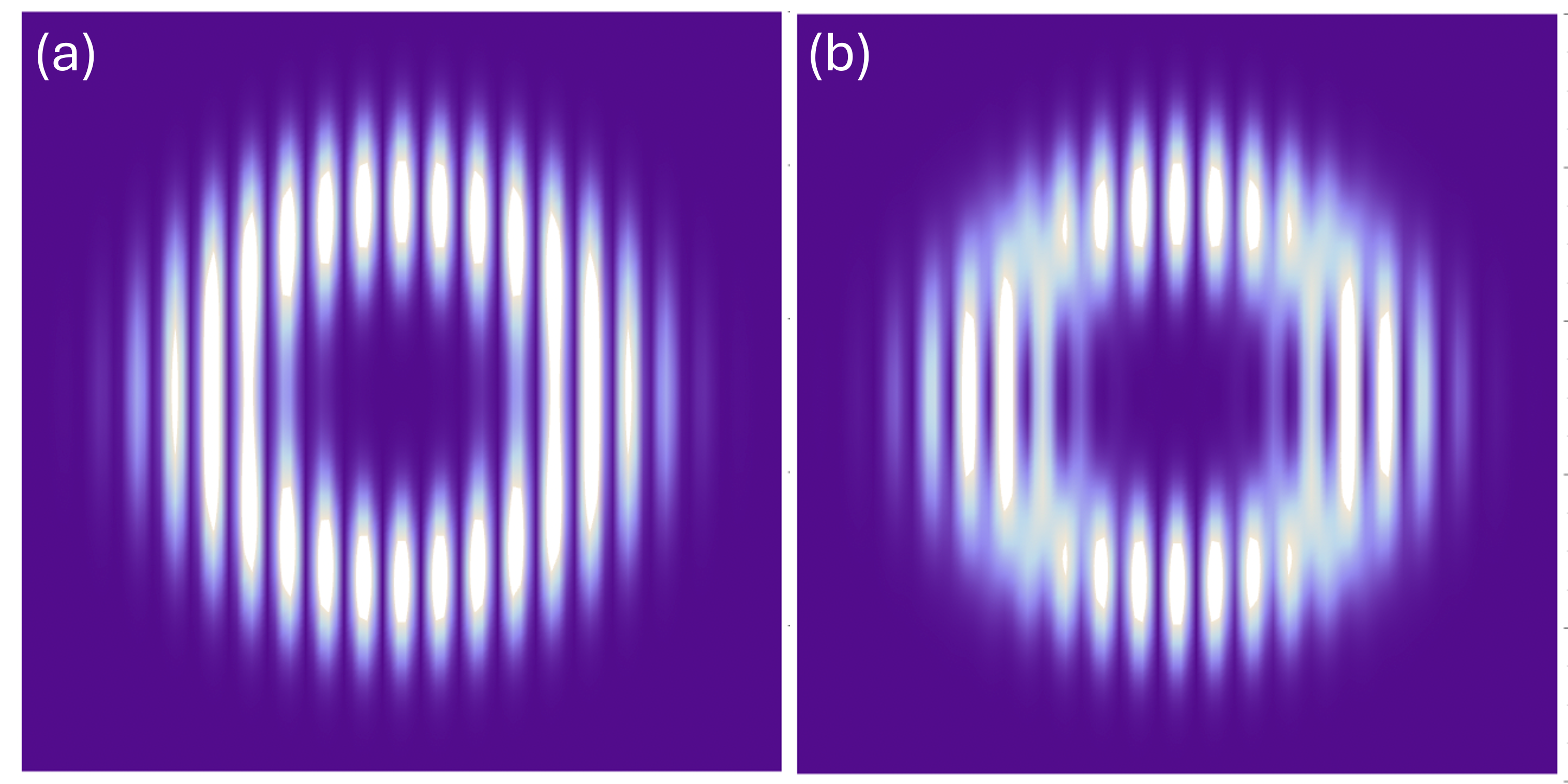}
    \caption{(a) interference pattern with no circulation of the condensate. (b) the sum of many interference patterns with an equal probability of being either left-handed circulation or right-handed circulation.} 
    \label{fringepatterns}
\end{figure}

The question may be raised whether the fringe patterns we see could be the result of the sum of a stochastic flipping between different circulations. In general, the answer is no, unless one type of circulation dominates the vast majority of instances. Figure \ref{fringepatterns} shows a comparison of a single interference pattern with no circulation of the condensate, and the pattern created by summing an equal number of patterns with left- and right-handed circulation. As seen in the image on the right side, a 50-50 stochastic mixture does not reproduce the interference pattern of a single condensate with no circulation.  There will always be blurring in some regions, determined by the alignment of the mirrors that give the two beams to create the fringes.

\section{Additional data of the experiment}
\subsection{Power series of the pump}
\begin{figure}[h]
    \centering
    \includegraphics[scale=0.6]{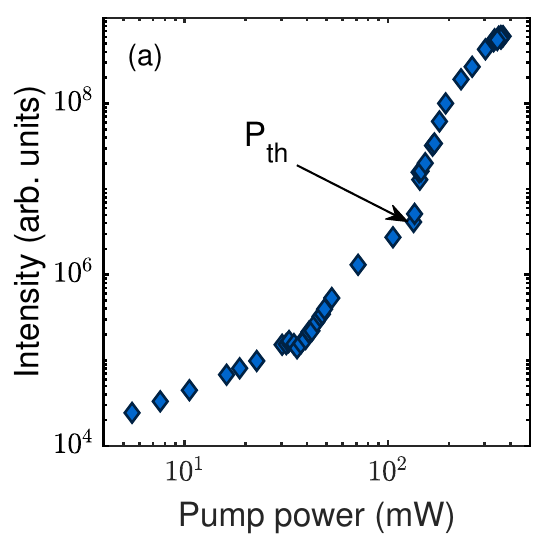}
    \includegraphics[scale=0.52]{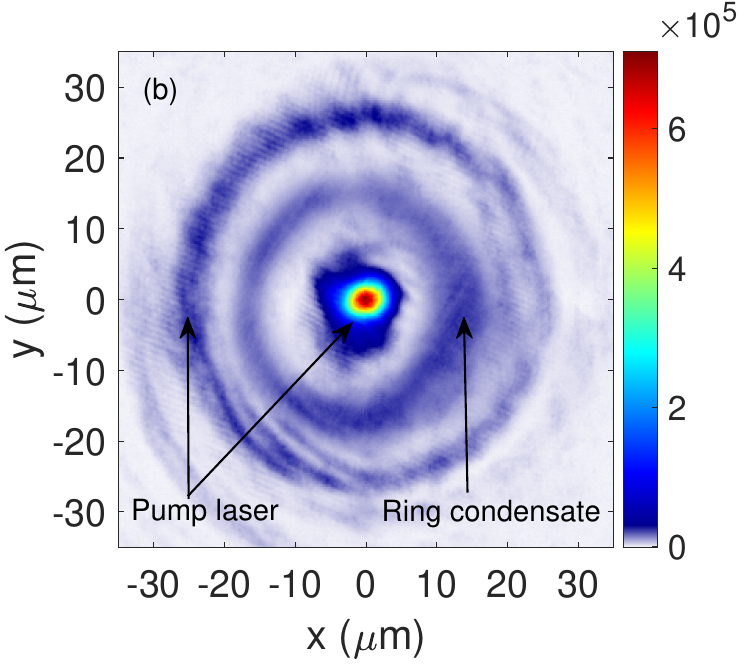}
    \caption{(a) Power series for the PL intensity of the condensate. (b) Image of the pump laser and polariton ring condensate at the highest pumping power. The image was taken without optical filters, so it captures signal from 400-900~nm. The pump laser wavelength is 730~nm and the condensate wavelength is 775~nm}
    \label{fig:pumpPseries}
\end{figure}

Fig.~\ref{fig:pumpPseries}(a) shows the total PL intensity of the ring versus the pump power. It indicates the threshold power was about 134~mW. In the experiment, an aperture was applied to avoid the higher-order patterns from SLM entering the sample and the power meter. 

Fig.~1(b) and Fig.1~(c) in the main text show the polariton photoluminescence (PL) below and above the threshold power. Usually, the pumping laser obstructs the observation of the PL due to its high brightness, even for a condensate. That's why a long-pass filter is always used before taking an image of the polariton PL. However, in our experiment, the polariton ring condensate forms in an optical trap, so we can image both the pumping laser and the PL, as shown in Fig.~\ref{fig:pumpPseries}(b). The central dot and the outer ring represent the target-shape pumping laser, while the ring between them corresponds to the polariton condensate.

\begin{figure}[h]
    \centering
    \includegraphics[scale=0.55]{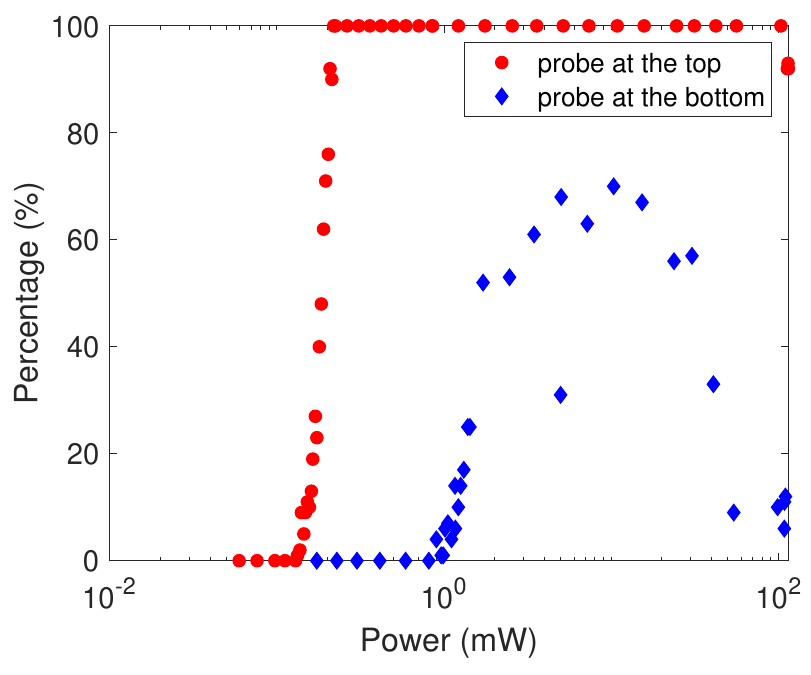}
    \caption{Percentage of circulation cases versus probe power. Red circles: probe was at the top of the ring. Blue diamonds: probe was at the bottom of the ring. The probe laser went through a filter wheel before reaching the sample so that we could change the power by rotating the filter wheel. However, the pitch and yaw angles of the filter wheel changes a little when rotating and the probe may not hit the ``good" position for starting circulation. This causes the percentage drop after $\sim11$mW in the blue data.}
    \label{fig:probPseries}
\end{figure}

\subsection{Power series of the probe}

In the experiment, we observed that there needs a minimum amount of power of the probe to make the circulation stable. In the experiment, we took 100 images of the interference pattern for each probe power, and counted the number of images which show circulation. Fig.~\ref{fig:probPseries} is the result. The data in red circles was taken when the probe was at the top of the ring, like the case in Fig. 2(a) in the main text. The data in blue diamonds was taken when the probe was at the bottom of the ring, like the case of Fig. 2(b) in the main text. It is easy to see that the threshold power for starting the circulation depends on the probe location. At some places, the probe couldn't cause circulation. We will discuss more about this in Section \ref{sec:8pos}. For the experiment in the main text, the probe power was $25.1$~mW, with $100\%$ showing circulation when the probe was at the top, while $85\%$ showing circulation when the probe was at the bottom.

\subsection{Polarization of the condensate}
\begin{figure}[h]
    \centering
    \includegraphics[scale=0.6]{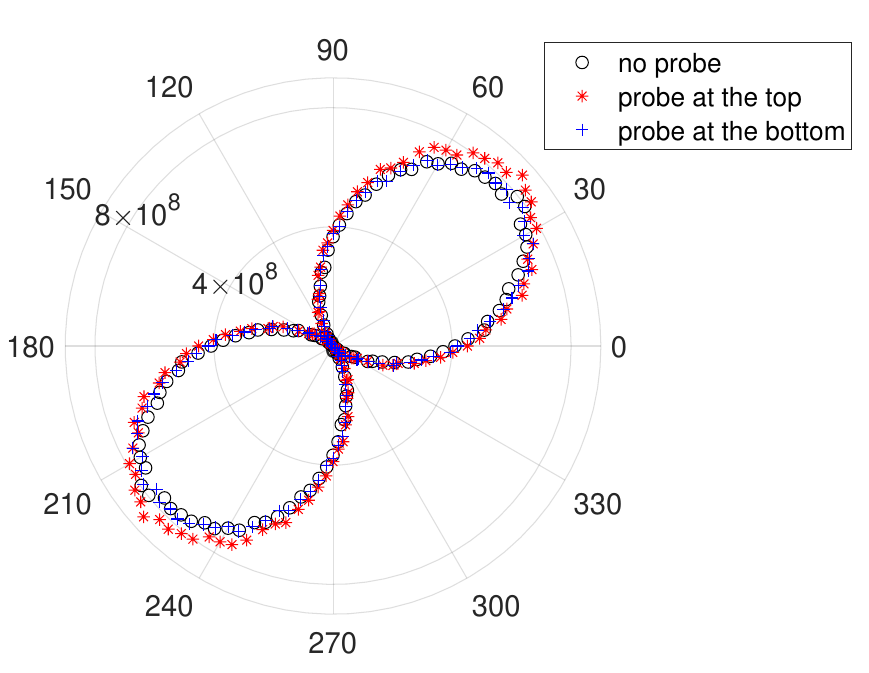}
    \caption{Intensity versus polarization. Data in black circle corresponds to Fig. 1(c) in the main text, when the condensate was not circulating. Data in red star corresponds to Fig. 2(a) in the main text, when the condensate was circulating. Data in blue cross corresponds to Fig. 2(b) in the main text, when the condensate was circulating in the other direction.}
    \label{fig:pol}
\end{figure}

\begin{figure}[h]
    \centering
    \includegraphics[scale=0.5]{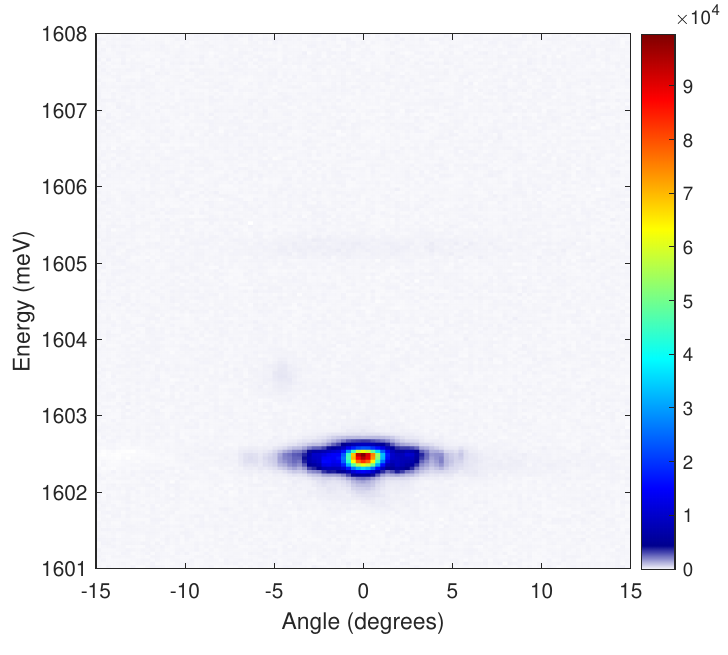}
    \caption{Dispersion of the ring condensate}
    \label{fig:k_PL}
\end{figure}

In Ref.~\cite{si10}, the spin procession of the condensate contributes to the circulation. In our case, the polarization of the condensate didn't change. Fig.~\ref{fig:pol} shows the intensity versus polarization in circulation and non-circulation cases. In order to minimize the role of TE-TM splitting, 
%we used a region of the sample with nearly zero gradient in the polariton energy and tune the pump laser profile very symmetric, 
we compensated the sample gradient by careful adjustment of the pump laser profile and made a trap with almost zero gradient,
so that the ring condensate wouldn't flow to any particular side and only occupied the $k=0$ state. Fig.~\ref{fig:k_PL} shows the momentum distribution of the ring condensate.

\newpage
\subsection{Moving the probe location}\label{sec:8pos}
\begin{figure}
    \centering
    \includegraphics[width=0.9\textwidth]{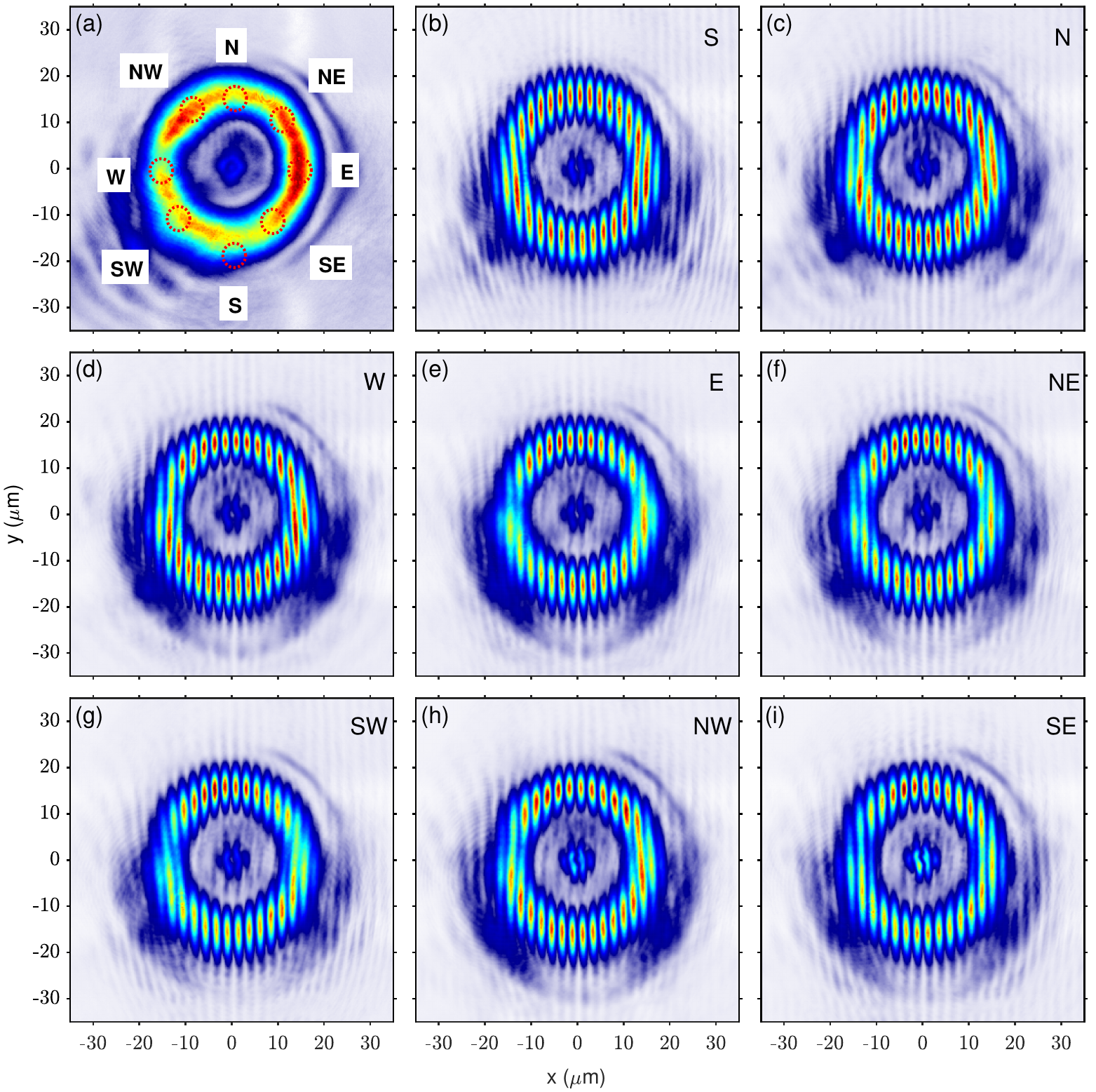}
    \caption{(a) Non-circulating ring condensate. Red circles label the probe position for (b)-(i). (b)-(i) interference patterns when introducing the probe laser. Letters on the top right indicate the probe location.}
    \label{fig:allPos}
\end{figure}

Fig. 2 in the main text shows the situation when the probe was at the top and bottom of the ring condensate. In fact, the probe doesn't need to be at exact ``top" or ``bottom" to make the condensate circulate. Fig.~\ref{fig:allPos} shows the ring condensate and eight interference patterns with the same probe power but at different probe positions. The intensity of the ring condensate, Fig.~\ref{fig:allPos}(a), is not perfectly uniform. When the probe was at the low intensity position of the ring, it could cause circulation, i.e.,  Fig.~\ref{fig:allPos}(b)(c)(d)(g)(h). When the probe was at the high intensity position of the ring, it couldn't cause circulation, i.e., Fig.~\ref{fig:allPos}(e)(f)(i).

The direction of the circulation was determined by the probe location and the intensity distribution of the ring condensate, which didn't change once the pump laser was set up. Each image in Fig.~\ref{fig:allPos} is an average of 100 subsequent images. The percentage showing circulation in Fig.~\ref{fig:allPos}(b)(c)(d)(g)(h) are $97\%$, $100\%$, $100\%$, $76\%$, and $98\%$, respectively. The percentage of not showing circulation in Fig.~\ref{fig:allPos}(e)(f)(i) are $88\%$, $100\%$, and $100\%$.

The circulation direction is more sensitive to the probe position when the probe moves in a region where the intensity of the ring condensate varies. For example, Fig.~\ref{fig:allPos}(d) shows circulation when the probe was on the west. If the probe was moved down a little to the high intensity region, the circulation disappeared. When the probe was moved to the southwest position, the circulation happened again with a different direction. On the contrast, the circulation direction is insensitive to the probe position when the probe moves in a uniform region, especially high intensity region, like from northeast to southeast, where no circulation was induced.

\subsection{Changing the symmetry of the probe}
\begin{figure}[t]
    \centering
    \includegraphics[scale = 0.4]{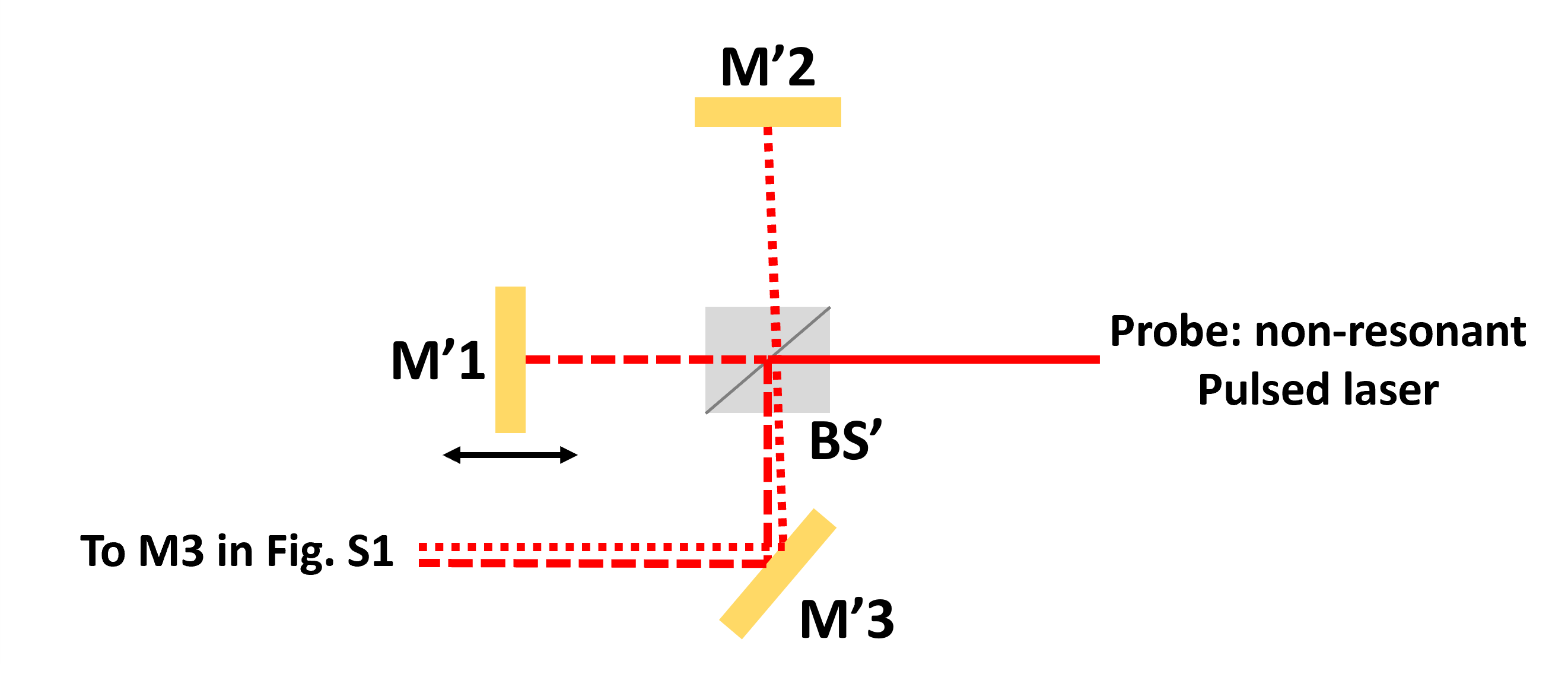}
    \caption{Schematic of the setup for changing the probe symmetry. BS: beam splitter; M: mirror (M$'$1 is on a translation stage)}
    \label{fig:probe_sym}
\end{figure}

The theoretical analysis shows that the asymmetry of the probe is essential to initialize the circulation. In the experiment, we tried to change the symmetry of the probe with the method proposed in Section~\ref{sec:theory}. Fig.~\ref{fig:probe_sym} is a schematic of the setup. The probe beam was split into two by the beam splitter BS$'$. Mirror M$'$1 and M$'$2 were of the same path length, and M$'$1 was on a translation stage. Different symmetry of the overlapped spot was obtained by moving one spot around the other. This was done by tilting M$'$2 and slightly translating M$'$1, which compensated the change of the path length. %The angle change by M$'$2 was very small and no interference pattern was observed, so the overlapped beam can be treated as one spot. 

However, the circulation was not affected in the experiments by our tuning of the degree of asymmetry of the probe. We believe this is because making a perfectly symmetric probe in reality is nearly impossible, as the beam goes through the lenses and the microscope objective (L1, L2 and MO in Fig.~\ref{fig:setup}). Because the system is quite sensitive to each small asymmetries, to give the overall symmetry breaking, only an extremely symmetric profile would not break the symmetry.

\subsection{Movies of the time-resolved interference patterns}

\begin{figure}[h]
    \centering
    \includegraphics[width=0.7\textwidth]{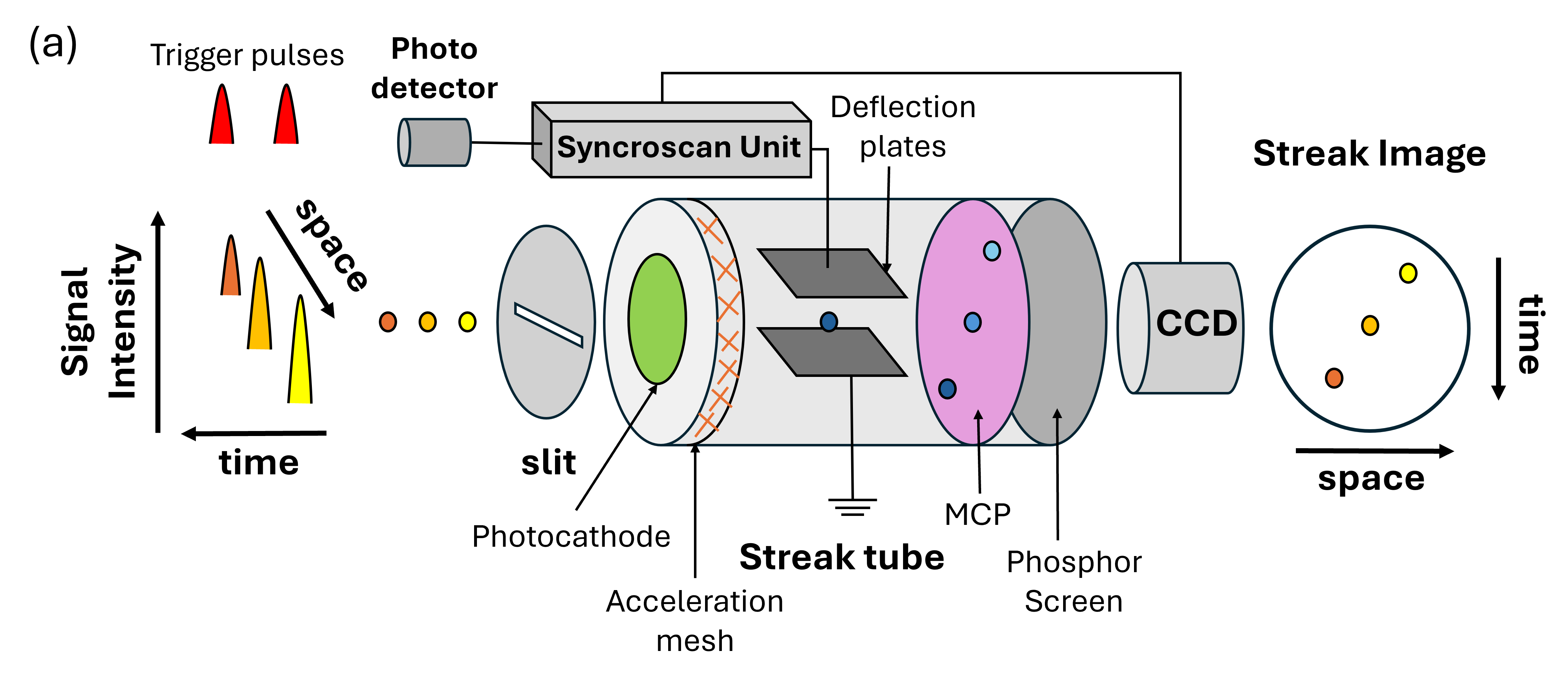}
    \includegraphics[scale=0.45]{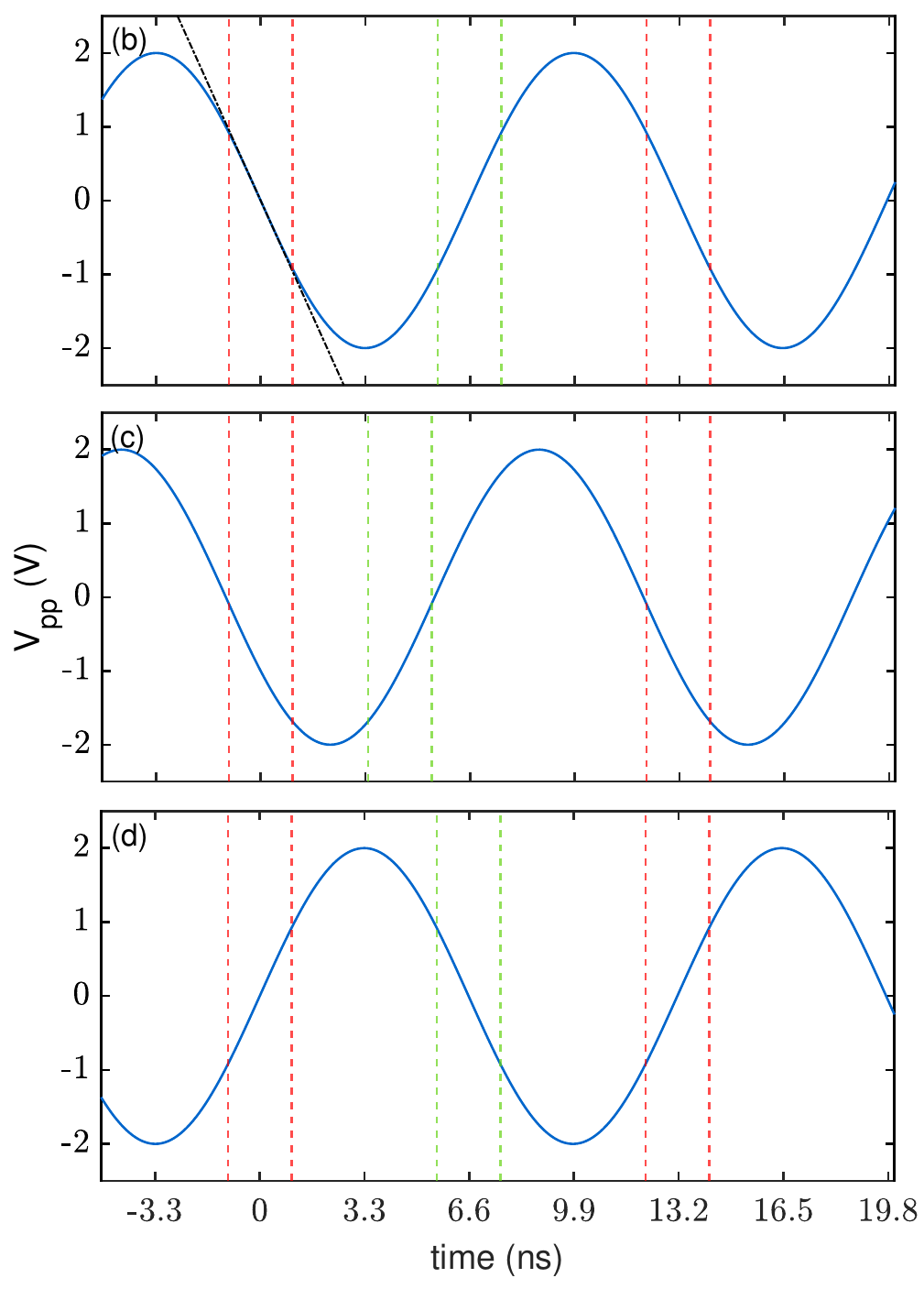}
    \caption{(a) Streak camera working principle, adapted from Ref.~\cite{streak-manual}. (b) correct phase delay of the sweeping voltage for measuring the signal during the time of interest, labeled by the red dash lines. The black dot line shows the linear part of the sweep. The reverse streak happens during the time between the green lines. (c) an incorrect setting of the phase delay. The time axis is no longer linear and the reverse streak is closer. (d) the phase delay is 180 degrees away from the correct setting.}
    \label{fig:streakRF}
\end{figure}

A Hamamatsu streak camera was used for taking the time-resolved images. Fig.~\ref{fig:streakRF}(a) shows the streak camera working principle. The camera first converts the light signal to electrons via a photocathode. Those electrons are accelerated and then fly through the streak tube, which contains a pair of metal plates, called deflection plates. The time-resolved imaging is realized by applying a sweeping voltage to the deflection plates, which gives the electrons a vertical displacement. Finally, the swept electrons are amplified by an microchannel plate (MCP) and converted back into light at the phosphor screen located at the end of the streak tube, so the signal can be detected by a CCD camera.

In our experiment, we use a synchroscan unit to provide the sweeping voltage, which is a sine wave with its phase and frequency locked to the pulsed laser repetition frequency (76MHz or 13.2ns). The phase delay of the sine-wave voltage is also tunable relative to the laser pulses, allowing the temporal measurement range to be within the linear region of the sine wave, ensuring a linear time axis. 

Fig.~\ref{fig:streakRF}(b)-(d) shows three examples of the phase delay setting, when we are interested in the signal from $t_1=-1000$~ps to $t_2=1000$~ps. Since the pulses enter the system every 13.2~ns, the final image is an average of signal between $(t_1+13.2\times q)$~ns and $(t_2+13.2\times q)$~ns, as $q=1,2,3,\cdots$, during the integration time. The time of interest is labeled by red dash lines. 
Fig.~\ref{fig:streakRF}(b) shows the correct setting of the phase delay. In this case, the linear region of the sine wave aligns with the time of the interest, and the electrons are swept from top to bottom. However, electrons generated during $(t_1+6.6)$~ns to $(t_2+6.6)$~ns feel the same voltage with an opposite sign, so they are swept from bottom to top and give an overlapping signal. This signal is called {\em reverse streak}. The green dash lines label the reverse streak time. If the phase delay is offset too much, shown in Fig.~\ref{fig:streakRF}(c), the time axis is no longer linear at the end of the sweep, and the reverse streak is closer. The streak camera is usually unstable in this situation, and will lose the phase lock frequently. Fig.~\ref{fig:streakRF}(d) shows when the phase delay is 180 degrees away from the correct setting. The streak camera could work stably this way but the time axis is reversed. We can identify this scenario by observing the polariton photoluminescence (PL) created by a non-resonant pulse.

A streak image only shows the time evolution of a slice of the entire image. We moved the final imaging lens, L5 in Fig.~\ref{fig:setup}, vertically to select different slices of the interference pattern, and took streak images for each slice with the same streak camera settings. As a result, by stitching the same row for each streak image, we obtain an image of the interference pattern at a particular time. A movie is made by playing those images in the order of their original row number. Fig.~3(a)-(c) in the main text are three frames of the full movie, Movie ``streak\_top.mp4". The movie for the situation when the probe shot at the bottom of the ring (Fig.~2(b) in the main text) is Movie ``streak\_bot.mp4". The time range of these two movies is 790~ps to have higher resolution in time. Fig.~3(e)(f) were extracted from streak images taken with a time range of 2168~ps. 

% streak\_top.mp4: full movie for Fig. 3 in the main text. This movie shows how the interference pattern changes when the probe shot at the top of the ring condensate.

% Streak\_bot.mp4: full movie of the interference pattern when the probe shot at the bottom of the ring condensate.

% \newpage
\section{Additional details of the theoretical model}\label{sec:theory}
The decay rates of condensed polaritons in Eqs. (2)-(3) of the main text was estimated as $\gamma_c = (1-\vert X\vert^2) \gamma_{ph}$, where $\gamma_{ph} = 1/\tau_{ph}$ corresponds to the inverse of the photon lifetime and $X$
is the Hopfield coefficient~\cite{yama-deng}. The excitonic fraction $|X|^2$ can be estimated as  $|X|^2~=~{1}/{2}\left(1+ {\delta}/{\sqrt{4 \hbar^2 \Omega^2 + \delta^2}} \right)$, where $\Omega$ is the Rabi splitting and the detuning $\delta$ corresponds to the difference between the photonic and excitonic energies. 

\begin{figure}[h]
    \centering
    \includegraphics[width=0.9\textwidth]{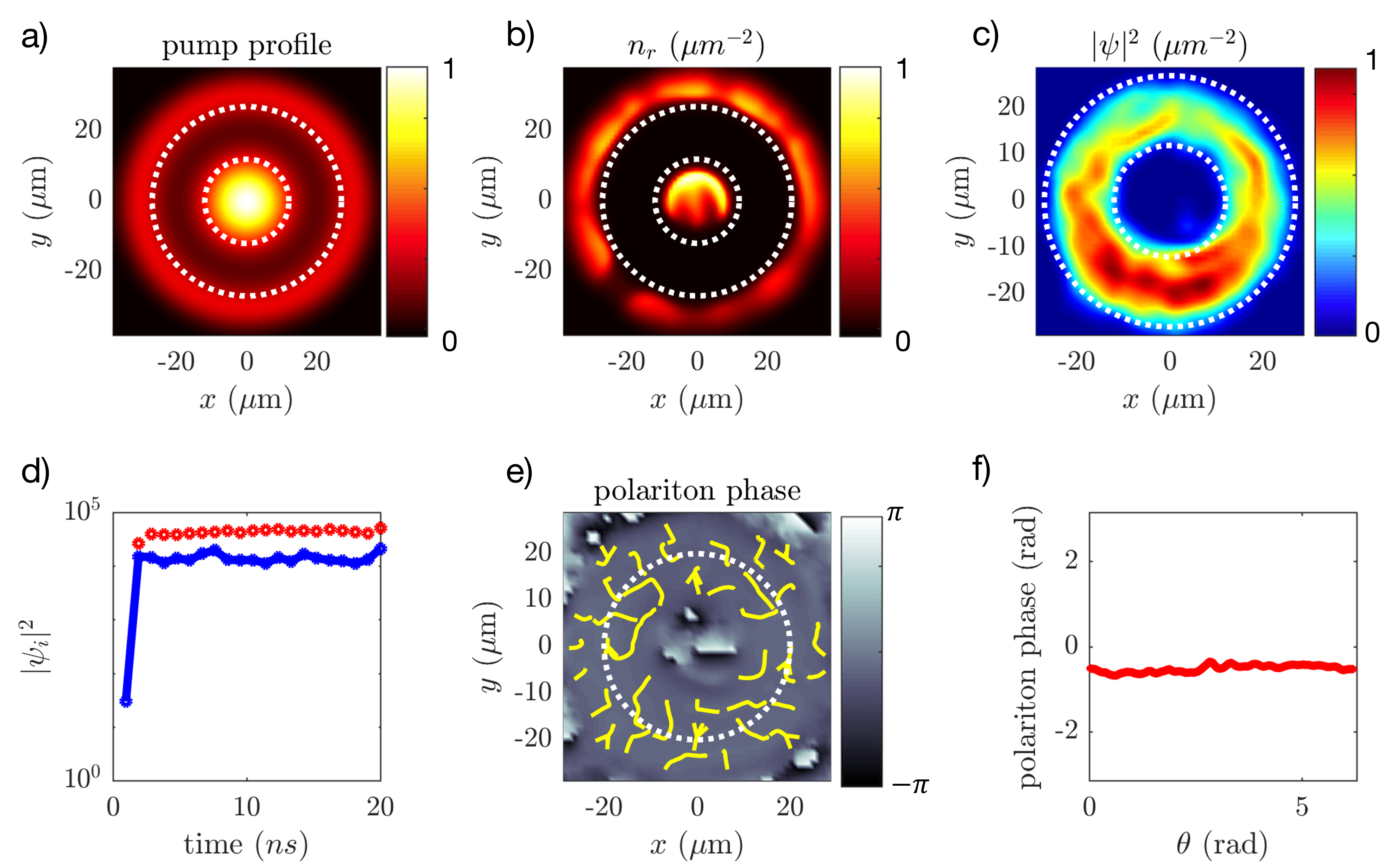}
    \caption{Numerical steady-state configuration. 
    (a) the pump profile, (b-c) the steady-state reservoir and polariton densities, (d) the total reservoir (red) and polariton density (blue) along the time evolution. (e) The polariton phase with streamlines, namely curves tangent to the local velocity vector (yellow lines) and (f) the phase along the central line of the toroidal potential (dashed white lines in (a)-(c)). All densities are normalised to the maximum values.} 
    \label{fig:simNocir}
\end{figure}

Fig.~\ref{fig:simNocir} shows the numerical steady-state configuration after running as single stochastic realisation of the theoretical model Eqs.~(2)-(3) of the main text, up to $t=0.1~\mathrm{\mu s}$. 
While Fig.~\ref{fig:simNocir}(a) show the two-dimensional projection of the pump profile, Fig.~\ref{fig:simNocir}(b) and Fig.~\ref{fig:simNocir}(c) show the normalised reservoir and polariton density at the end of the stochastic evolution.
The temporal evolution of the normalised density of the two field is depicted in Fig.~\ref{fig:simNocir}(d).
The two-dimensional phase field and the unwrapped phase profile are shown in Fig.~\ref{fig:simNocir}(e)-(f), respectively.
Depending on the initial noise seed chosen ($dW_c$ in Eq. (2) of the main text), the steady-state can have different spatial profiles.
This is due to the hole-burning effect~\cite{estrecho2018single} introduced by the repulsive exciton-polariton interactions. We use this effect to imitate a spatially modulated long-range disorder.
To be consistent with the experimental observations, (Fig.~2 in the main text), in our analysis we choose a particular stochastic realization whose polariton density had a well-defined minimum (``valley'') and maximum (``hill'') in the ring (Fig.~\ref{fig:simNocir}(c)).

In our numerical model many properties of the probe could be varied, namely its position, its intensity, and its spatial profile. 
As already introduced in the main text, the probe is modelled as an external exciton-reservoir potential density $n_R^\mathrm{probe} = H_{R} \mathcal{P} (1+\chi dW(t))$, where the term $H_{R}$ corresponds to its intensity. The probe spatial profile $\mathcal{P}$ is written as the sum of two different overlapping functions, $\mathcal{P}(\textbf{r}_1,\textbf{r}_2) = \mathcal{G}_1(\textbf{r}_1)+\mathcal{G}_2(\textbf{r}_2)$, which allowed us to explore the effect of the probe asymmetry on the imprinting of the circular currents. 
We also model classical random laser fluctuations by means of the term $dW(t)$, namely a real-valued zero-mean, Gaussian noise.
For simplicity, we modeled the spatial asymmetry of the probe pulse as two adjacent Gaussian functions $\mathcal{G}_1(\textbf{r}_1)$ and $\mathcal{G}_2(\textbf{r}_2)$, located at the position $r_1$, $r_2$ respectively. As a result, the probe asymmetry was controlled by changing the position and the intensity of $\mathcal{G}_1$ and $\mathcal{G}_2$. To mimic a pulse of the laser, the probe was switched on for a period of $T_\mathrm{probe} = 10$~ps, one order of magnitude smaller than the polariton lifetime. We then let the system evolve and investigated its dynamics.

We repeated this numerical simulation varying many parameters of the pumping, such as the driving strength $P$, the ratio $H$, random small fluctuations of the spatial potential-energy profile, as well as parameters of the short probe pulse: for a symmetric probe, we can tune its amplitude and RMS width, while in the case of an asymmetric probe, we could vary the ratio between the intensities of $\mathcal{G}_1$ and $\mathcal{G}_2$, and their positions. We found that the effect of circulation
induced by the probe is robust under a wide variety of conditions, but
requires the probe to be at least slightly asymmetric.

The model was numerically solved by adopting an explicit Runge-Kutta method of orders eight and nine with fixed time-step in a two-dimensional numerical grid with $N=128^2$ points with grid-spacing $a = 1.17 \si{\mu m}$. 
In order to match experimental observations, we solved the dynamical equations with the following parameters: $m=4.2 10^{-5} m_e$ with $m_e$ the electron mass, $\tau_{ph} = 135 \si{ps}$, $\gamma_R = 10^{-3} \mathrm{ps}^{-1}$, $g_c~=~2~\si{\mu eV \mu m^2}$, $R_0 = 2 \times 10^{-3} \si{\mu m^2 ps^{-1}}$. 
The parameters of the pump rings read $\sigma_{ring}= 3.72 \si{\mu m}$, $\sigma_{gauss}= 8.43 \si{\mu m}$, $R_{ring} =30 \si{\mu m}$, while the amplitude ratio between the central pump and the ring is~$3$.
The probe intensity $H_R$ is tuned such that it matches the reservoir ring maximum, while the intensity of the real potential reads $\chi =0.1$.
The two Gaussian functions forming the probe $\mathcal{P}$ have the same standard deviation $\sigma_{\mathcal{G}_1} = \sigma_{\mathcal{G}_2} = 4 \mathrm{\mu m}$ but different intensities; $4 \ 10^3 \mathrm{\mu m^{-2}}$ and $2 \ 10^3 \mathrm{\mu m^{-2}}$ respectively. 

% \FloatBarrier

\section{Review of Earlier Work on Persistent Circulation}

As discussed in the introduction to the main text, several papers have been published which discuss persistent circulation of polariton condensates. While each of these represented significant progress in the field, technically, none of them can be described as a true example of persistent circulation. In this section, we give some additional information on these papers.

In early work \cite{si5}, quantized circulation was seen in time-averaged interference imaging over a region about 2 microns in size in a disordered landscape. 
No time-resolved data were taken. The quantized circulation was visible in time-averaged imaging because some vortices created in a highly-nonequilibrium condensate became pinned at disorder minima. A later, similar experiment \cite{si6} showed the same type of behavior with ``half-vortices'' (quantized circulation with both current flow and polarization rotation). In both of these experiments, the flow could not be stopped or reversed in the location observed. The most likely explanation for the circulation was that the disorder produced a ``toilet-bowl'' static potential profile with built-in chirality, which amounted to a type of constant stirring. 

In Reference \cite{si7}, quantized circulation was observed in a similar far-from-equilibrium polariton experiment, without steady state, with time resolution over a range of about 120 ps.  The time-resolved data showed a decay time of the vortex of about 100 ps.  This was longer than the time that the laser was on, so that it may be correctly said that the vortex persisted longer than it was stirred, but is not an example of persistent current, as the decay time was finite (equivalent to nonzero resistivity in a metallic conductor). Similar work \cite{si7,si8} using time-resolved correlation-function imaging measured a coherence decay time of about 100 ps, not persistence.

\begin{figure}[t]
    \centering
    \includegraphics[scale = 0.45]{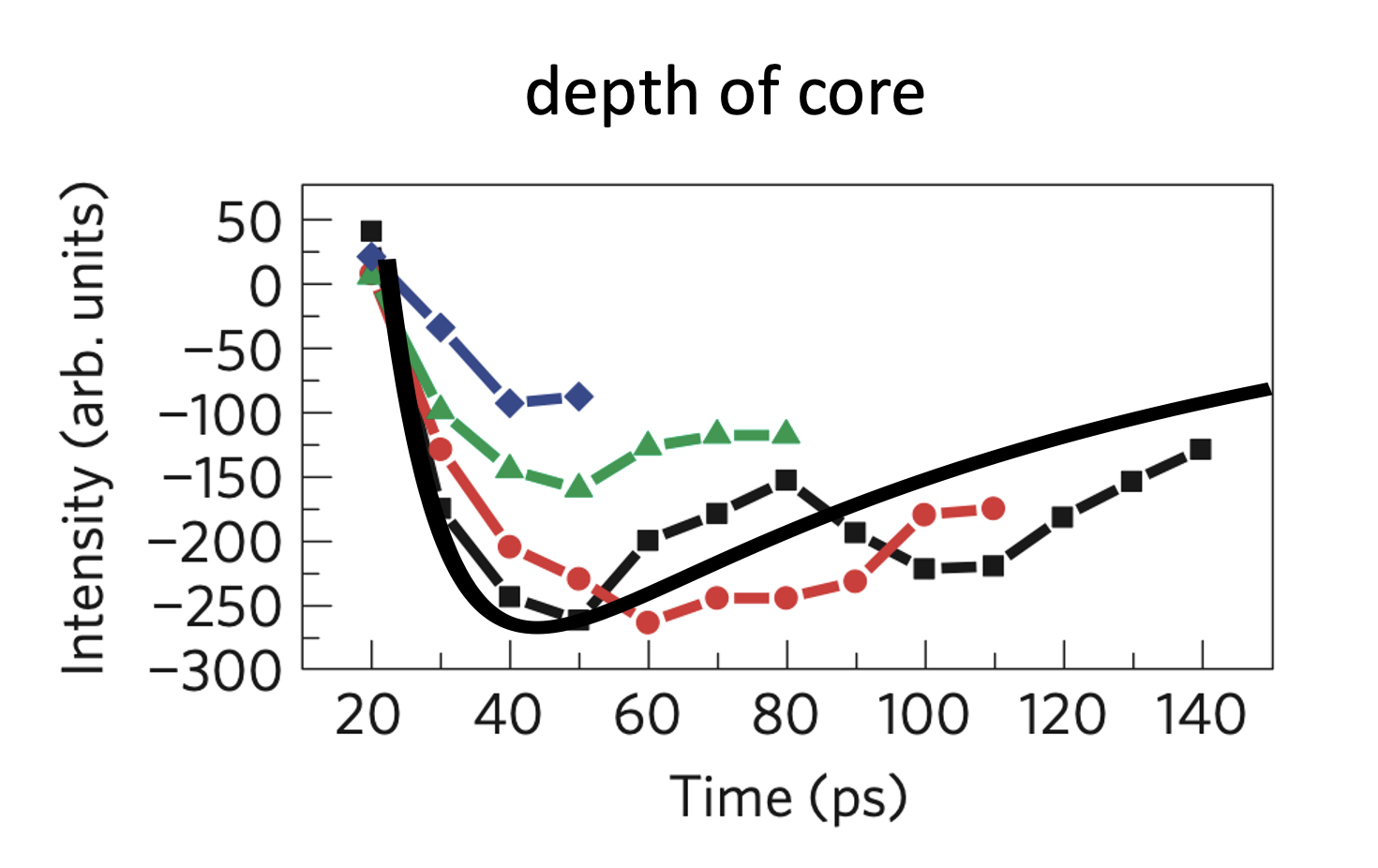}
    \caption{Data points: relative change of the vortex core intensity as a function of time, for four pump powers, reproduced from Ref.~\protect\cite{si8}.  Solid black line: A simple time-dependent curve with 10 ps onset time and 100 ps decay time.} 
    \label{SI_sanvitto_fig}
\end{figure}

In Ref.~\cite{si9}, time-resolved measurements were reported which tracked, for about 30 ps, the coherence of a half-vortex induced by a pulsed laser about 2 microns in size. As reported in that paper, the vortices were not stable, but split on time scales of 10 ps.  Theoretical modeling indicated that the instability was due to disorder in the system. 

Ref.~\cite{si12} reported a coherent standing wave of a condensate in steady state. The standing wave of the condensate was pinned by self-trapping, via a mechanism that is well known to create a ``beaded'' pattern of the condensate with periodic maxima \cite{si13}. The coherent standing wave in Ref.~\cite{si12} was interpreted as a superposition of two counter-propagating circulating condensates, since a $k$-space analysis showed two peaks with opposite momenta, but this amounts to a mathematical Fourier analysis. As is well known in Fourier mathematics, any standing wave  can be viewed as a superposition of two counter-propagating waves with opposite $\vec{k}$. Since no actual current was measured in these experiments, it is more proper to say that this experiment showed long-range coherence, but no persistent circulation.  A similar experiment was done in Ref.~\cite{si14}, which showed long-range coherence of standing waves in a Mach-Zehnder geometry. 

In Refs.~\cite{si15} and \cite{si16}, steady-state circulation of a polariton was seen in systems in which a driving force caused the circulation. No persistent circulation was reported when the driving force was turned off. These are therefore not examples of the effect of persistent circulation.

% In Refs.~\cite{si15} and \cite{si16}, steady-state circulation of a polariton was seen in systems in which a driving force caused the circulation. No persistent circulation was reported when the driving force was turned off.

In earlier work including some of us~\cite{si10}, a circulating ring condensate was created in steady state, with coherence length of more than 40 microns, which had time intervals of stability but stochastic switching between the circulation direction on long time scales. Crucially, this earlier work could not show stable {\em lack} of circulation, and the direction of circulation could not be controlled, which left open the interpretation that the circulation could be due to a metastable chirality in the disorder landscape, as in Refs.~\cite{si5,si6}. Ref.~\cite{si11} recently showed a result very similar to that of Ref.~\cite{si10}. 

In Ref.~\cite{si17}, the phase winding of a condensate was not measured directly, and instead a Fourier transform of the spatial pattern was used. The direction of circulation flipped stochastically, and so could not be called persistent, but a predominance of one direction of circulation over the other was found.

In contrast to all of the above, the experiments reported here show a truly persistent, steady-state circulation with no measurable dissipation (analogous to zero resistance of a superconductor). As seen in Figure~3 of the main text, the phase winding measured from the interference patterns is absolutely stable after the initial transients for as long as we can measure. The fringe visibility also shows no trend of decay over the same time period, within the noise. 

%As discussed in Section \ref{sec:8pos}, since this is an ensemble-averaged measurement, and the fringe visibility is not 100\%, we cannot absolutely rule out the possibility that in some instances, there is no circulation. However, as discussed in that section, the stability of the phase winding implies that the great majority of the instances do indeed persist in circulation with the same handedness for the whole time; if there was stochastic switching of the circulation direction, as in Refs.~\cite{si10} and \cite{si11}, we would see blurring in some parts of the images, or multiple images with different circulation directions for the same conditions, as was observed in Ref.~\cite{si10}. 

Finally, we discuss some experiments in atomic and optical physics. In a purely optical experiment \cite{si18}, two specific modes of an optical resonator could be selected so that their interference pattern had a time-varying oscillation that corresponded to circulation. This experiment is in many ways analogous to Refs.~\cite{si15} and \cite{si16}, in that a driving force was used to created constant circulation. 

Ref.~\cite{si19} detected the presence of a vortex of cold atoms as a function of time by measuring the density profile at various times and looking for the vortex core. The phase winding of the condensate was not measured, but the presence of the core showed no decay in time above a critical density. Ref.~\cite{si20} did a similar measurement through the BEC-BCS crossover. 

A later paper \cite{si21} measured the phase winding of an atomic condensate in using atom interferometry, without showing time-resolved measurements.  More recently, Ref.~\cite{si22} measured the phase winding of an atomic condensate in the BCS regime using atom interferometry, with time resolution. The current persisted well past the stirring force but showed decay over time, primarily because the number of atoms in the trap decreased over time.  There has been as of yet no successful true steady-state condensate of atoms that can be examined by interferometry {\em in situ}. This is one of the appeals of the polariton microcavity system, that the dynamics of a condensate including its many-body wave function can be studied in equilibrium and steady state, as only a tiny fraction of the photons escape and are replenished at any point in time.

\bibliography{ref_si}